\documentclass[preprint]{revtex4-1}

\usepackage{hyperref}
\usepackage{graphicx}
\usepackage{amstext}
\usepackage{amsmath}
\usepackage{amssymb} 
\usepackage{cleveref}
\usepackage{color}
\usepackage{subcaption}
\usepackage{array}

\begin{document}

\def\ra{\rangle}
\def\la{\langle}
\def\nn{\nonumber}
\def\dsp{\displaystyle}
\def\mc{\mathcal}
\def\pr{\prime}
\def\sc{Schr\"{o}dinger }
\def\ep{\epsilon}
\def\h{\hat}
\def\HF{\textrm{u}}
\def\pd{\partial}
\def\te{\text}
\def\da{\dagger}
\def\cua{\h{d}^{\dagger}_{1\uparrow}}
\def\cda{\h{d}^{\dagger}_{1\downarrow}}
\def\aua{\h{d}_{1\uparrow}}
\def\ada{\h{d}_{1\downarrow}}
\def\cub{\h{d}^{\dagger}_{2\uparrow}}
\def\cdb{\h{d}^{\dagger}_{2\downarrow}}
\def\aub{\h{d}_{2\uparrow}}
\def\adb{\h{d}_{2\downarrow}}
\newcommand{\bra}[1]{\langle #1|}
\newcommand{\ket}[1]{| #1\rangle}

\newcommand{\fig}[1]{Fig.~\ref{#1}}
\newcommand{\Fig}[1]{Figure~\ref{#1}}
\newcommand{\eq}[1]{Eq.~(\ref{#1})}
\newcommand{\Eq}[1]{Equation~(\ref{#1})}

\title{Understanding the Fundamental Connection Between Electronic Correlation and Decoherence}
\author{Arnab Kar}
\author{Liping Chen}
\author{Ignacio Franco}
\email{ignacio.franco@rochester.edu}
\affiliation{Department of Chemistry and The Center for Coherence and Quantum Optics, University of Rochester, Rochester, New York 14627, USA}



\begin{abstract}
We introduce a theory that exposes the fundamental and previously overlooked connection between the correlation among electrons and the degree of quantum coherence of electronic states in matter. For arbitrary states, the effects only decouple when the electronic dynamics induced by the nuclear bath is pure-dephasing in nature such that  $[H^{S},H^{SB}]=0$, where $H^{S}$ is the electronic Hamiltonian and $H^{SB}$ is the electron-nuclear coupling. We quantitatively illustrate this connection via exact simulations of a Hubbard-Holstein molecule using the Hierarchical Equations of Motion that show that increasing the degree of electronic interactions can enhance or suppress the rate of electronic coherence loss.

Published in: A.~Kar, L.~Chen, and I.~Franco, J. Phys. Chem. Lett. \textbf{7}, 1616 (2016).
\end{abstract}

\maketitle

Understanding the behavior of electrons in matter is fundamental to our ability to characterize, design and control the properties of molecules and materials~\cite{Stefanucci,Nitzan}. Electronic correlations~\cite{Szabo,Fetter} and decoherence~\cite{Breuer,Schlosshauer,Joos} are two basic properties that are ubiquitously used to characterize the nature and quality of electronic quantum states. Correlations among electrons arise due to their pairwise Coulombic interactions, that lead to a dependency of the motion of an electron with that of other surrounding electrons.  These correlations determine the energetic properties of electrons in matter and the character of their energy eigenstates~\cite{Kohn,Pople}. In turn, decoherence in molecules typically arises due to the interactions of the electrons with the nuclear degrees of freedom~\cite{Hwang,Prezhdo_QuantumDecoherence,Franco_Decoherencedynamics}. The nuclei act as an environment that induces a loss of phase relationship between quantum electronic states. Establishing mechanisms for electronic decoherence  is central to our understanding of the excited state dynamics of molecules~\cite{photo1,photo2,photo3, scholes2015}, to the development of useful approximations to model correlated electron-nuclear  dynamics~\cite{Kapral,Prezdho_DISH}, and to the design of strategies to preserve electronic coherence that can subsequently be  exploited in quantum technologies~\cite{QuantumControl_Shapiro, Nielsen}. 

While electronic correlation and decoherence have been amply investigated separately, { the connection between the two, if any, is not  understood. This is  partially due to the fact that usual} definitions of electronic correlation, such as correlation energy~\cite{Lowdin} or natural occupation numbers~\cite{Ziesche}, are only applicable to pure electronic systems~\cite{Eberly1,Nest} and do not allow addressing this fundamental question.  For this reason,  it is unclear if decoherence can induce changes in  correlation and, conversely, if  correlations can modify the coherence content of a quantum state. 
 
Here we demonstrate that electronic correlation and decoherence are coupled physical phenomena that need to be considered concurrently. We do so by extending the concept of electronic correlation to open non-equilibrium quantum systems, and showing that electronic correlation modulates the degree of entanglement between electrons and nuclei, and thus the degree of electronic decoherence. Conversely, we also show that the electronic decoherence modulates the degree of electronic correlation, as evidenced by the correlation energy.  Further, we isolate conditions under which electronic correlations and decoherence can be considered as uncoupled physical phenomena and show that they are generally violated by molecules and materials, demonstrating that the connection between electronic correlation and decoherence is ubiquitous in matter. These formal developments are quantitatively illustrated via numerically exact computations in a Hubbard-Holstein molecule that show that increasing the electronic interactions can strongly modulate the rate of electronic coherence loss.

To proceed, consider a pure electron-nuclear system with Hamiltonian 
$\mc{H}=H^S+H^B+H^{SB}$,
where $H^S$ is the electronic Hamiltonian, $H^B$ the nuclear component, and $H^{SB}$ the electron-nuclear couplings.  {Here, $H^{SB}$ is defined as the residual electron-nuclear interactions that arise when the nuclear geometry deviates from a given reference configuration (e.g., the optimal geometry).} The electronic Hamiltonian $H^S=H^S_0+V^S$ can be further decomposed into single-particle contributions $H^S_0$ (e.g. Hartree-Fock) and { residual} two-body terms $V^S$. The latter arise from Coulombic interactions { that cannot be mapped into one-body terms} and introduce correlations among the electrons. The associated non-interacting Hamiltonian is obtained when $V^S=0$, and is given by $\mc{H}_0=H^S_0+H^B+H^{SB}$.

To extend the concept of electronic correlations to open non-equilibrium  quantum systems, we require a  correlation metric  and a reference uncorrelated state for each electron-nuclear state. To construct the reference state, we imagine a fictitious process where for each physical time $t$ the  $V^S$  term in the Hamiltonian is turned off adiabatically slow along a fictitious time coordinate $\tau$  (see Fig. S1 in the Supporting Information (SI)). Specifically, we suppose that the Hamiltonian of the system is of the form 
\begin{align}
\label{GML}
\mc{H}_{\epsilon}(\tau)=\mc{H}-e^{-\epsilon|\tau|}V^S \quad (\ep>0),
\end{align}
where the second term is considered as a perturbation to the $\mc{H}$-induced evolution. The physical evolution along $t$ occurs at the $\tau = \tau_0\to-\infty$ limit of the $(t, \tau)$ space for which the Hamiltonian is in its fully interacting form $\mc{H}_{\epsilon}(\tau_0)=\mc{H}$. In this limit, the state of the fully interacting system is given by
\begin{align}
\hat{\rho}(t)&=\dsp{\sum_{i,j}\alpha_i(t)\alpha_j^*(t)|\psi_i\ra\la\psi_j|},\label{initial}
\end{align}
where $\ket{\psi_i}$ are eigenstates of $\mc{H}$ ($\mc{H}\ket{\psi_i} = E_i \ket{\psi_i}$). The uncorrelated reference state is generated by adiabatically turning off, in the Interaction picture, the $V^S$ term in the Hamiltonian in the $\tau=\tau_0$ to $\tau=0$ interval, i.e.
\begin{align}
\label{UtoHFinter}
\hat{\rho}^{\HF}(t)&=\lim_{\ep\to 0} \lim_{\tau_0\to -\infty} U_{\ep I}(0,\tau_0)\hat{\rho}(t)U_{\ep I}^\dagger(0,\tau_0),
\end{align}
where $U_{\epsilon I}(\tau,\tau^{\prime})$ is the evolution operator in Interaction picture. The latter is defined by the  Dyson series~\cite{Fetter} $U_{\epsilon I}(\tau,\tau^{\prime})=\mathbb{I}+\sum_{n=1}^{\infty} U^{(n)}_{\epsilon I}(\tau,\tau^{\prime})$, where
\begin{align}
U^{(n)}_{\epsilon I}(\tau,\tau^{\prime})=-\frac{i}{\hbar}\displaystyle{\int_{\tau^{\prime}}^{\tau}d\tau_n e^{-\epsilon |\tau_n|} V_{I}(\tau_n)U^{(n-1)}_{\epsilon I}(\tau_{n},\tau^{\prime})}\nn,
\end{align}
$V_I(\tau)=- U_0^\dagger(\tau)V^S U_0(\tau)$ is the $-V^{S}$ operator in  Interaction picture, and $U_0(\tau)= e^{-\frac{i}{\hbar}\mc{H}\tau}$ is the perturbation-free evolution operator.

\Eq{UtoHFinter} captures  changes in $\hat{\rho}(t)$ that are generated by the process of turning off  $V^S$ in the presence of a nuclear environment.  It has the desirable property that $\hat{\rho}^\HF(t)= \hat{\rho}(t)$ when $V^S=0$, and it reduces to the usual adiabatic connection for isolated electronic systems when {$H^{SB}=0$}.  Note that we have chosen $U_{\epsilon I}(\tau)$ instead of the full evolution operator $U(\tau) =U_0(\tau) U_{\epsilon I}(\tau)$ to generate the uncorrelated states. This is because the $U_0(\tau)$ component of $U(\tau)$ leads to  changes in $\hat{\rho}(t)$ due to  electron-nuclear entanglements that are present even when $V^{S}=0$. By contrast, $U_{\epsilon I}(\tau)$ solely captures electron-nuclear entanglements that can be  modulated by the  electron-electron interactions.

Switching off  interactions adiabatically  generates exact eigenstates of the non-interacting system from those of the interacting system via the Gell-Mann and Low theorem (GMLT)~\cite{Fetter,Stefanucci}. The GMLT states that given an eigenstate $\ket{\psi_i}$ of the interacting $\mc{H}$, if the limit 
\begin{align}
\label{evolve}
\lim_{\epsilon \to 0}|\phi^{\ep}_{i}\rangle=\lim_{\ep \to 0}A_i^{-1} U_{\epsilon I}(0, -\infty)|\psi_i\ra,
\end{align}
(where $A_i=\bra{\psi_i} U_{\ep I}(0,-\infty) \ket{\psi_i}/|\bra{\psi_i} U_{\ep I}(0,-\infty) \ket{\psi_i}|$ and $|A_i|^2=1$ because the $|\phi_i\ra$ are chosen to be normalized) exists, then $\lim_{\ep \to 0}|\phi^{\ep}_i\ra=|\phi_i\ra$ is an eigenstate of the non-interacting $\mc{H}_0$. Applying the GMLT in \eq{UtoHFinter} we arrive at the  uncorrelated reference state that corresponds to $\h{\rho}(t)$ in \eq{initial}, 
\begin{align}
\hat{\rho}^{\HF}(t)
=\dsp{\sum_{i,j} \alpha_i(t)\alpha_j^*(t)e^{i(\theta_i-\theta_j)}|\phi_{i}\ra\la\phi_j|}.\label{limitHF}
\end{align}
Here, we have assumed that $\lim_{\epsilon \to 0}\,A_iA_j^\star=e^{i(\theta_i-\theta_j)}$ exists even when the phase factors $A_i \sim e^{\frac{i}{\epsilon}}$ are known to be ill-behaved as $\epsilon\to 0$~\cite{Fetter}. While the $A_i$ introduce convergence issues at the wavefunction level, observable quantities, including the density operator, should remain finite during the unitary evolution. 

As a physical measure of electronic correlation in electron-nuclear systems we choose the energetic difference between the correlated and uncorrelated state:
\begin{align}
E_{\text{cor}}(t)&=\text{Tr}[\hat{\rho}(t)\mc{H}]-\text{Tr}[\hat{\rho}^{\HF}(t) \mc{H}_0]\label{ecor}.
\end{align}
This quantity measures energetic changes in the electron-nuclear system that are introduced by the process of turning off {$V^S$}  during the adiabatic connection in \eq{UtoHFinter}, and parallels a common metric for correlation~\cite{Lowdin} used in closed electronic systems. Note that any energetic measure of  correlation based on the properties of the electronic subsystem alone is not appropriate since it will unavoidably include relaxation channels due to interactions with the bath. Further note that definitions of  correlation based on the non-idempotency of the single-particle electronic density matrix~\cite{Ziesche, Mukherjee} are not applicable since the non-idempotency can arise due to correlation or  due to decoherence~\cite{Franco_Reduced}.
 (see Ref.~\cite{Mauser, GottliebJQI}  for measures claimed to operate in open quantum systems).

As a basis-independent measure of decoherence we employ the purity $P(t)=\text{Tr}[\hat{\rho}_e^2(t)]$ where $\h{\rho}_e(t)=\text{Tr}_B[\h{\rho}(t)]$ is the $N$-body electronic density matrix obtained by performing a partial trace over the nuclear  bath. The purity $P=1$  for pure states and $P<1$ for mixed states.  For pure electron-nuclear systems, the decoherence of the  electronic (or nuclear) subsystem is solely due to electron-nuclear entanglement. Thus, in this regime, the decay of $P$ also measures the degree of electron-nuclear entanglement.

In this context, it is now readily seen why  correlation and decoherence are strongly connected. For this, first note that the coherence content of $\hat{\rho}_e(t)$ and $\hat{\rho}_e^{\HF}(t)$ are generally different. To see this, consider  $\hat{\rho}(t)=|\Psi(t)\ra\la\Psi(t)|$ in \eq{initial} for which $|\Psi(t)\ra=\sum_i  \alpha_i(t)|\psi_i\ra$. In light of the Schmidt decomposition~\cite{Nielsen}, $|\Psi(t)\ra$ can  be written as $|\Psi(t)\rangle=\sum_{i} \sqrt{\lambda_i(t)}|s_i(t)\rangle |b_i(t)\ra$, where $|s_i(t)\ra$ and $|b_i(t)\ra$ are, respectively, orthonormal electron and nuclear states, and $\sqrt{\lambda_i}$ are the Schmidt coefficients $(\sum_{i} \lambda_i=1, \lambda_i>0)$. In the Schmidt basis,
\begin{align}
\h{\rho}(t)=\dsp{\sum_{i,j}\sqrt{\lambda_i\lambda_j}|s_i\ra|b_i\ra\la b_j|\la s_j|}\label{schmidt}.
\end{align}
In terms of $\{\lambda_i\}$, the purity of the electronic (or nuclear) subsystem is $P(t)=\sum_i \lambda^2_i(t)$.
In turn, the uncorrelated state $\h{\rho}^{\HF}(t)=|\Phi(t)\ra\la\Phi(t)|$ (\eq{limitHF}) is associated with $|\Phi(t)\ra\equiv\sum_{i} \alpha_i(t)e^{i\theta_i}|\phi_{i}\ra$. Under the Schmidt decomposition, $|\Phi(t)\ra=\sum_i \sqrt{\mu_i(t)}|S_i(t)\ra|B_i(t)\ra$ and the resulting purity is $P^{\HF}(t)=\sum_i \mu_i^2(t)$. 
Since $|\Phi(t)\ra\neq |\Psi(t)\ra$, the set $\{\mu_i\}$ is different from the set $\{\lambda_i\}$ and therefore the purity for the correlated state and its reference uncorrelated counterpart generally differ.  That is, for ${H^{SB}}\ne 0$,  $V^S$ modulates  the degree of coherence of electronic states.

Consider now the influence of $H^{SB}$  on the correlation energy [\eq{ecor}],
\begin{equation}
E_{\text{cor}}(t)=\sum_i |\alpha_i(t)|^2 (E_i-\mc{E}_i),\label{ecorex}
\end{equation}
where $\mc{H}|\psi_i\ra=E_i|\psi_i\ra$ and $\mc{H}_0|\phi_i\ra=\mc{E}_i|\phi_i\ra$. For $V^S\neq 0$, $E_\text{cor}$ will change if  $H^{SB}$ changes because $E_i$ and  $\mc{E}_i$ vary differently as $H^{SB}$ is modified.  That is, $H^{SB}$ influences $E_{\text{cor}}$ because it modulates the response of the electron-nuclear system to $V^S$. 

Decoherence and correlation  decouple when 
\begin{align}
\label{conditions} 
[H^S,H^{SB}]=0,
\end{align}
for $V^S\ne 0$. When \eq{conditions} holds, the $H^{SB}$ does not introduce electronic relaxation  and the system-bath dynamics is pure dephasing. To see how this sufficient condition arises, consider the decoherence case first. For the purity of $\h{\rho}_e(t)$ and $\h{\rho}_e^{\HF}(t)$ to coincide, the evolution operator in \eq{UtoHFinter} must not change the degree of entanglement between electrons and nuclei. For this to happen, $U_{\ep I}$ must be of the form
\begin{align}
U_{\epsilon I}(0,-\infty)=U^S_{\epsilon I}(0,-\infty)\otimes\mathbb{I}^B,\label{U}
\end{align}
where $U^S_{\epsilon I}$ is a purely electronic operator and $\mathbb{I}^B$ is the identity operator in the nuclear Hilbert space. Under these conditions, and in light of \eq{schmidt},
$
\hat{\rho}^{\HF}(t)=\sum_{i,j} \sqrt{\lambda_i\lambda_j} |s^{\pr}_i\ra|b_i\ra\otimes\la b_j|\la s^{\pr}_j|,
$
where $ \ket{s_i'}\bra{s_j'}=\lim_{\ep \to 0}U^S_{\epsilon I}(0,-\infty)\ket{s_i} \bra{s_j}{U^S_{\epsilon I}}^\dagger(0,-\infty)$. Since the Schmidt coefficients for $\h{\rho}^{\HF}(t)$ are the same as those of $\h{\rho}(t)$ (\emph{cf.} \eq{schmidt}) the purity of the two states is identical. For $U_{\epsilon I}$ to be of the form in \eq{U}, $V_I(\tau)$ must be a purely electronic operator, i.e. $V_I(\tau)=\hat{O}^S(\tau)\otimes\mathbb{I}^B$, where $\hat{O}^S$ is an operator in the Hilbert space of the electronic subsystem. This is guaranteed when  \eq{conditions} is satisfied. Specifically, 
\begin{align}
V_I(\tau)&=-e^{i\frac{\tau}{\hbar}H^S}e^{i\frac{\tau}{\hbar}(H^B+H^{SB})}V^Se^{-i\frac{\tau}{\hbar}(H^B+H^{SB})}e^{-i\frac{\tau}{\hbar}H^S},\nn
\end{align}
where we have used the fact that  $[H^S,H^B]=0$  and the condition in \eq{conditions}.  We arrive at the desired form 
\begin{equation}\label{VI}
V_I(\tau) = -e^{i\frac{\tau}{\hbar}H^S}\!V^S\!e^{-i\frac{\tau}{\hbar}H^S}\!\otimes\!\mathbb{I}^B\!=\!\hat{O}^S\!(\tau)\!\otimes\!\mathbb{I}^B,
\end{equation}
by taking into account that $[V^S,H^B]=0$, and the fact that $[V^S,H^{SB}]=0$ for Coulombic systems since $V^S$ and $H^{SB}$ are both functions of the position operators.

The correlation energy also becomes independent of $H^{SB}$  when the  commutation relations in \eq{conditions} are satisfied. To show this, we contrast $E_{\te{cor}}$  with the correlation energy $E_{\te{cor}}^{(0)}$ that would have been obtained if $H^{SB}$ is not allowed to influence the response of the system as $V^S$ is adiabatically turned off  in \eq{UtoHFinter}. Specifically, 
\begin{align}
\label{ecor0}
E^{(0)}_{\te{cor}}(t)&=\text{Tr}[\h{\rho}(t)\mc{H}]-\text{Tr}[\h{\rho}^{\HF}_{(0)}(t)\mc{H}_0],
\end{align}
where the reference state $\hat{\rho}^{\HF}_{(0)}(t)=\lim_{ \tau_0\to -\infty, \ep \to 0}\left[U'_{\ep I}(0, \tau_0)\hat{\rho}(t)U_{\ep I}^{\pr \dagger}(0, \tau_0)\right]$ is obtained by setting $H^{SB}=\!0$ throughout the adiabatic process, i.e.  $U'_{\ep I}(0, -\infty)=U_{\ep I}(0,-\infty)\vert_{H^{SB}=0}$.
The interaction potential $V_I^{\pr}(\tau)$ in $U^{\pr}_{\ep I}(t,-\infty)$ is given by
\begin{equation}\label{hsbref}
V^{\pr}_I(\tau)=V_{I}(\tau) \vert_{H^{SB}=0}=-e^{i\frac{\tau}{\hbar}H^S}V^Se^{-i\frac{\tau}{\hbar}H^S}\otimes\mathbb{I}^B
\end{equation}
where we have used the fact that $[H^S,H^B]=[V^S,H^B]=\!0$. If $E_{\te{cor}}(t)=E_{\te{cor}}^{(0)}(t)$ the correlation energy is independent of $H^{SB}$. For this to happen, the identity  $V_{I}(\tau)=V^{\pr}_I(\tau)$ must be satisfied such that $\h{\rho}^{\HF}_{(0)}(t)$ and $\h{\rho}^{\HF}(t)$ coincide. Since $V_I'(\tau)$ is identical to the limiting $V_I(\tau)$ in \eq{VI},  by the same argument employed to arrive at \eq{VI} we conclude that $E_{\te{cor}}(t)=E_{\te{cor}}^{(0)}(t)$  when \eq{conditions} is true.

From the perspective of the correlation energy,  when \eq{conditions} is satisfied $E_{\te{cor}}$ is purely determined by the electronic subsystem. This is because $\la H^{SB}+H^B\ra$ remains constant as $V^S$ is turned off adiabatically (as can be seen by writing the Heisenberg equations of motion for $H^{SB}+H^{B}$). From the perspective of the purity, \eq{conditions}  guarantees that the effect of the bath will be the same for the correlated system and its uncorrelated counterpart, thus eliminating a possible $V^S$ dependence in the decoherence dynamics. 
 Note that even for stationary Born-Oppenheimer (BO) states it is not possible for decoherence and correlation to be uncoupled unless \eq{conditions} is satisfied. This is because even when stationary BO states are not entangled, the corresponding uncorrelated state generally will be. 

The pure dephasing condition [\eq{conditions}] is generally violated by molecules and materials, indicating that the connection between electronic correlation and decoherence is ubiquitous in matter. Nevertheless, pure dephasing dynamics can arise when the frequencies associated with nuclear motion are far detuned from the electronic transitions such that the nuclear dynamics does not lead to electronic transitions in the correlated and uncorrelated system, as can be the case in semiconducting quantum dots~\cite{QD1, QD2}.  Under such conditions, $H^{SB} \approx \sum_n\hat{F_n}\otimes \ket{E_n}\bra{E_n}$, where $\{\ket{E_n}\}$ are the eigenstates of $H^S$, and the $\hat{F_n}$ are nuclear operators defined such that $[V^S, H^{SB}]=0$.

We now quantitatively illustrate this connection using a neutral two-site, two-electron, Hubbard-Holstein model  with zero net spin as an example~\cite{Stefanucci}; a minimal molecular model that violates the commutation relations in \eq{conditions} and satisfies $[V^S, H^{SB}]=0$ as is expected for molecules. 
Here the electrons are described by the Hubbard Hamiltonian
\begin{equation}
H^S=-t_0\sum_{\sigma\in\{\uparrow,\downarrow\}}(\h{d}^{\dagger}_{1\sigma}\h{d}_{2\sigma}+\h{d}^{\dagger}_{2\sigma}\h{d}_{1\sigma})+U(\h{n}_{1\uparrow}\h{n}_{1\downarrow}+\h{n}_{2\uparrow}\h{n}_{2\downarrow})
\end{equation}
where $\h{d}^{\da}_{i\sigma}$ (or $\h{d}_{i\sigma}$) creates (or annihilates) an electron on site $i$ with spin $\sigma$  and satisfies the usual anticommutation relations $\{\h{d}_{i\sigma}, \h{d}^{\da}_{j\sigma'}\}=\delta_{i,j}\delta_{\sigma,\sigma^{\pr}}$. The quantity $\h{n}_{i\sigma}=\h{d}^{\da}_{i\sigma}\h{d}_{i\sigma}$ is the number operator, $t_0$ is the hopping parameter, and $U$ is the energy penalty for having two electrons on the same site.  
The Hubbard Hamiltonian can be decomposed into a Hartree-Fock component
$H^S_0=-t_0\sum_{\sigma\in\{\uparrow,\downarrow\}}(\h{d}^{\dagger}_{1\sigma}\h{d}_{2\sigma}+\h{d}^{\dagger}_{2\sigma}\h{d}_{1\sigma})+2U\sum_{i,\sigma}\h{n}_{i\sigma}\la \h{n}_{i,-\sigma}\ra-U\sum_{i,\sigma}\la \h{n}_{i\sigma}\ra\la \h{n}_{i,-\sigma}\ra$, and a two-body term $V^S=H^S-H^S_0$, where the expectation value $\la \h{n}_{i\sigma}\ra=1/2$  is over the equilibrium thermal state. 
The nuclei are described as four  baths of $N_b^m$ harmonic oscillators, with Hamiltonian
\begin{equation}
H^B=\sum_{m=1}^4\sum_{j=1}^{N_b^m}\left(\frac{p_{mj}^2}{2}+\frac{1}{2}\omega^2_{mj}x^2_{mj}\right),
\end{equation}
where  $x _{mj}$ is the mass-weighted displacement away from equilibrium for the $j$th harmonic oscillator  in the $m$th harmonic bath, $p_{mj}$ is the momentum conjugate to $x_{mj}$ and $\omega_{mj}$ its oscillation frequency.
We assume that each set of harmonic oscillators couples to an independent electronic  configuration of zero net spin. Specifically, we choose
\begin{equation}
\label{eq:hsb}
H^{SB}=F_1\h{n}_{1\uparrow}\h{n}_{1\downarrow}+F_2\h{n}_{2\uparrow}\h{n}_{2\downarrow}+F_3\h{n}_{1\uparrow}\h{n}_{2\downarrow}+F_4\h{n}_{1\downarrow}\h{n}_{2\uparrow},
\end{equation}
where $F_m=\sum_{j=1}^{N_b^m}c_{mj}x_{mj}$ is a collective bath coordinate of bath $m$. The effective electron-nuclear coupling is specified by the spectral density 
$J_m(\omega)=\frac{\pi}{2}\sum_{j=1}^{N_b^m} \frac{c_{mj}^2}{\omega_{mj}}\delta(\omega-\omega_{mj})$ which is assumed to be same for all the states and of Debye form 
$J(\omega)=\eta \frac{\gamma \omega}{\omega^2+\gamma^2}$.
Here $\gamma$ is the characteristic frequency of the bath and the parameter $\eta$ effectively determines the electron-nuclear coupling strength.

The electronic dynamics generated by this  model is propagated exactly using the Hierarchical Equations of Motion approach~\cite{Tanimura,Liping,HEOM1,HEOM2}, a non-perturbative and non-Markovian theory of reduced system dynamics.  As an initial state, we consider a separable electron-nuclear state $\hat{\rho}(0)= \hat{\rho}_e(0)\otimes \hat{\rho}_n(0)$ where the nuclei are initially at thermal equilibrium $\hat{\rho}_n(0)= \exp(-\beta H^{B})/\textrm{Tr}_B\{  \exp(-\beta H^{B}) \}$ with inverse temperature $\beta$, and the electrons $\hat{\rho}_e(0) = \ket{\Psi}\bra{\Psi}$  in a superposition  $\ket{\Psi}=\frac{1}{\sqrt{2}}(|E_1\ra+|E_2\ra)$ between the ground   and first excited   state.

\begin{figure}[htbp]
\includegraphics[width=0.5\textwidth]{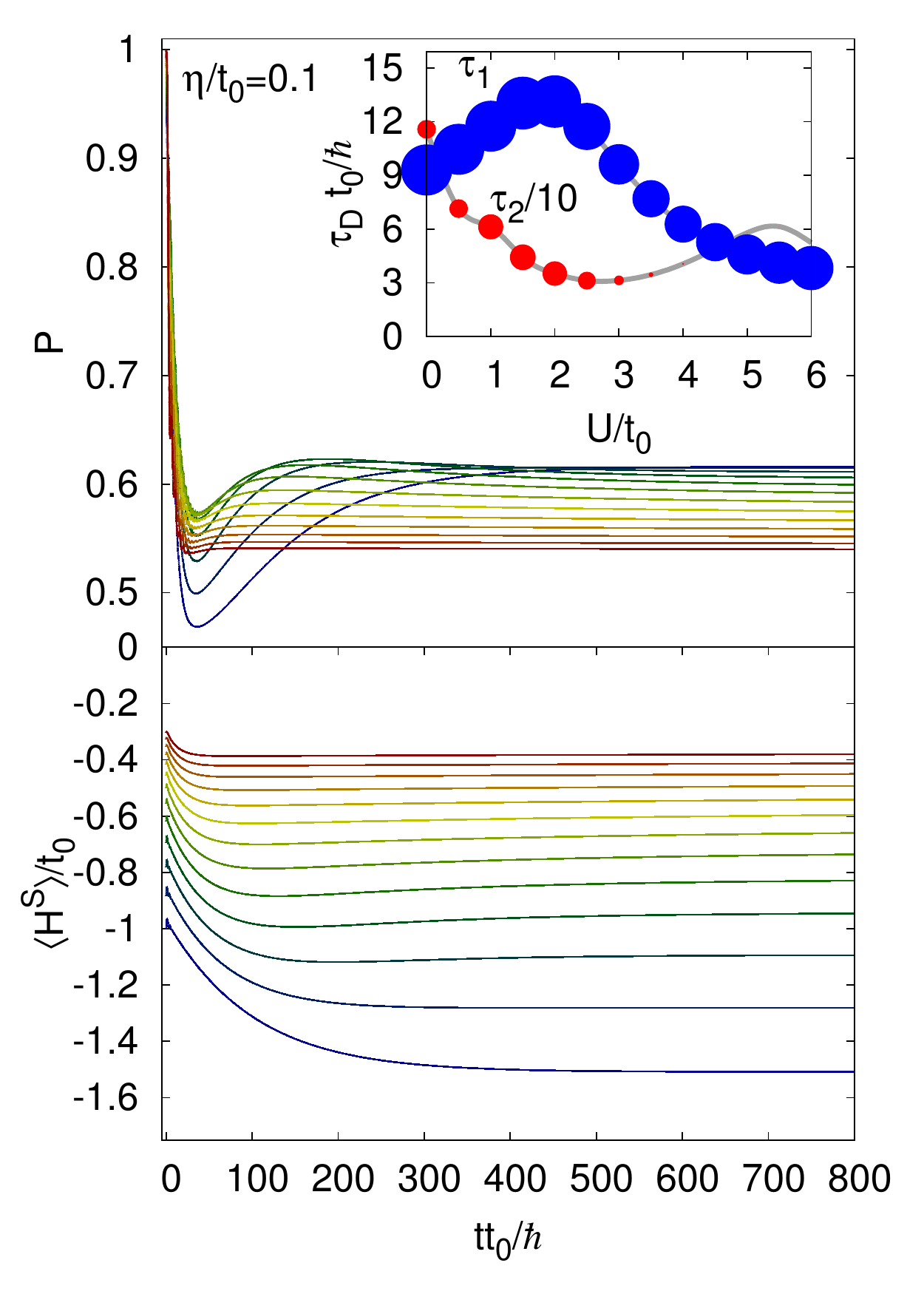}
\hspace{-0.22 in}
\includegraphics[width=0.5\textwidth]{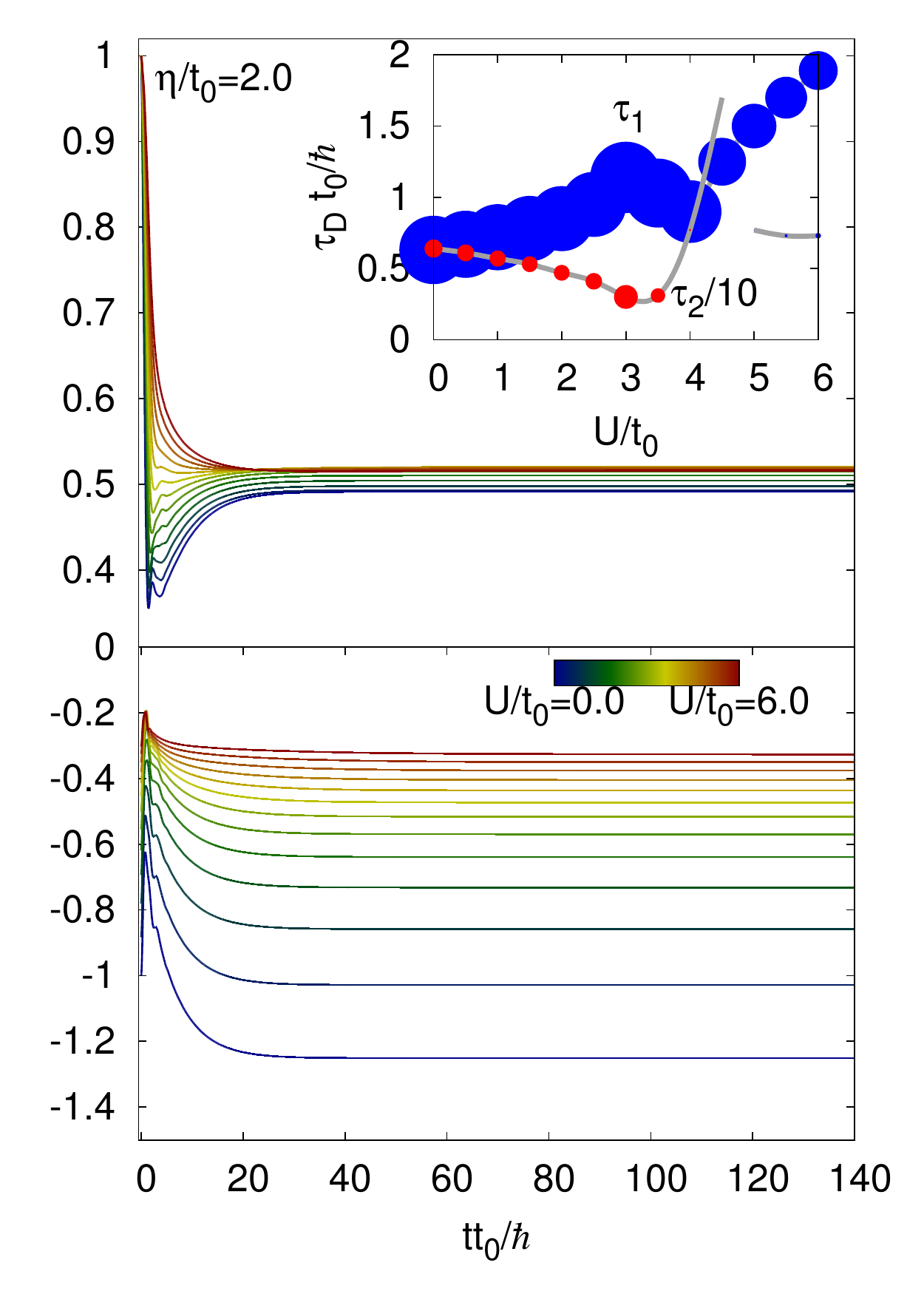}
\caption{\raggedright Purity and electronic energy during the evolution of the Hubbard-Holstein molecule ($\beta=1/t_0, { \hbar} \gamma=0.3 t_0$). 
The inset shows characteristic timescales in $P(t)$ obtained from an exponential fit $P-P_{\text{thermal}}=\sum_{i=1}^3 a_i \exp(-t/\tau_i)$, see SI. The dot size measures the magnitude of $|a_i|$ (blue, $a_i> 0$; red,  $a_i<0$; $a_3$ is small and not shown). Note how increasing $U$ can enhance  or suppress the decoherence.}\label{dynamics}
\end{figure}

\Fig{dynamics} shows the dynamics of the purity and the electronic energy for different electronic interactions $U$ and effective electron-nuclear couplings $\eta$ (inset: characteristic  decay timescales $\tau_i$ in $P(t)$). The fact that \eq{conditions} is violated is reflected by the  energetic relaxation of the electrons. The purity observes a sharp initial decay on a $\tau_1$ timescale, followed by a slower dynamics  on a $\tau_2$ timescale that asymptotically leads the electronic subsystem to a state of thermal equilibrium. In the presence of electronic correlations, varying $U$ strongly modulates the decoherence and relaxation dynamics. By contrast, in the Hartree-Fock approximation  the purity for this model is independent of $U$ and equal to the one for $U=0$. For $U\le 3t_0$ the decoherence is determined by $\tau_2$. In turn, for $U\ge 4t_0$ the importance of $\tau_2$ in the dynamics (as characterized by the dot sizes in \fig{dynamics}) is diminished,  and the decoherence time is determined by $\tau_1$. \emph{Note how increasing $U$ can enhance or suppress the rate of electronic decoherence.} Specifically, for $\eta=0.1t_0$, increasing $U$ leads to a  decrease in the decoherence time.  By contrast, for $\eta=2.0t_0$,  increasing $U$ leads to a decrease followed by an increase in the  decoherence time. As expected, the rate of decoherence is  faster in the stronger $\eta$ case.

The molecular mechanisms at play in \fig{dynamics} can be identified by examining the effect of changing $\eta$ and $U$ on the  potential energy surfaces (PESs). As detailed in the SI,  increasing $U$ brings the ground and first excited state closer together in energy, and  reduces the difference in curvature between their PESs. The first effect increases the decoherence rate  because it increases the nonadiabatic couplings between the two  states. Excitation by an incoherent bath leads to decoherence~\cite{vision}. Thus, the enhanced excitation of the electrons by the  \emph{thermal} nuclei  increases the  decoherence rate.  The second effect, by contrast, slows down the decoherence. To see this, recall that for a general vibronic state $\ket{\Psi} = \sum_n \ket{E_n} \ket{\chi_n}$ the electronic density matrix is given by $\hat{\rho}_e= \sum_{nm} \bra{\chi_m} \chi_n\rangle \ket{E_n}\bra{E_m}$. The coherences between states $\ket{E_n}$ and $\ket{E_m}$ are thus  determined by the nuclear wavepacket overlap $S_{mn}= \bra{\chi_m} \chi_n\rangle$~\cite{Prezdho_QuantumTransitionRates,Franco_Decoherencedynamics}. By making the PESs look more alike, increasing $U$ slows down the  decay of $S_{mn}$ for each member of the initial ensemble due to wavepacket evolution in alternative PESs.  It is the non-trivial competition between these two effects what leads to the intricate dynamics in \fig{dynamics}. 

Note that the nonadiabatic couplings between the ground (singlet) and first excited (triplet) state that are responsible for the first decoherence mechanism arise due to the $F_3\h{n}_{1\uparrow}\h{n}_{2\downarrow}$ and $F_4\h{n}_{1\downarrow}\h{n}_{2\uparrow}$ terms in $H^{SB}$. By contrast, the second decoherence mechanism is determined by all four terms in $H^{SB}$ and survives even in the absence of singlet-triplet couplings. In this limit, increasing $U$ protects the electrons from the decoherence. 

\begin{figure}[htbp]
\includegraphics[width=0.5\textwidth]{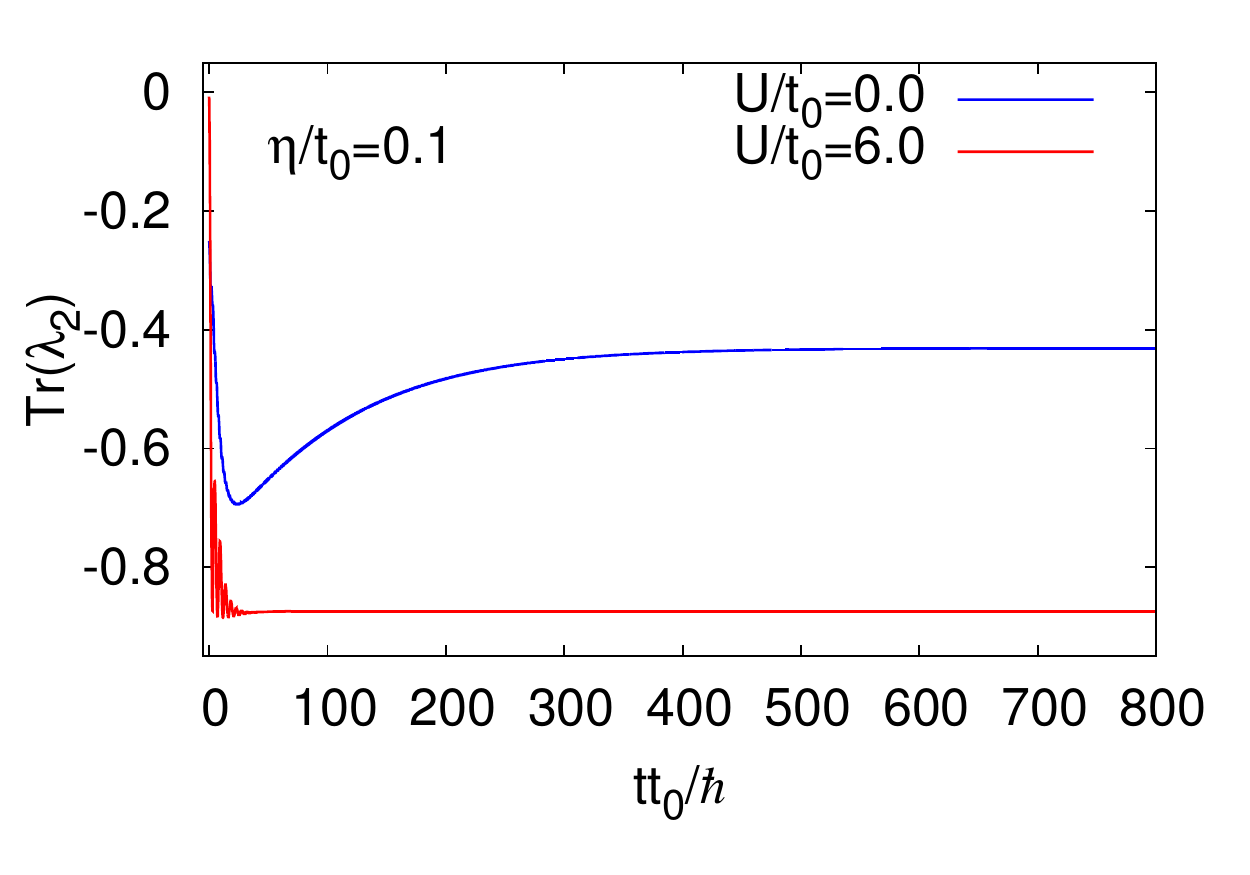}
\hspace{-0.22 in}
\includegraphics[width=0.5\textwidth]{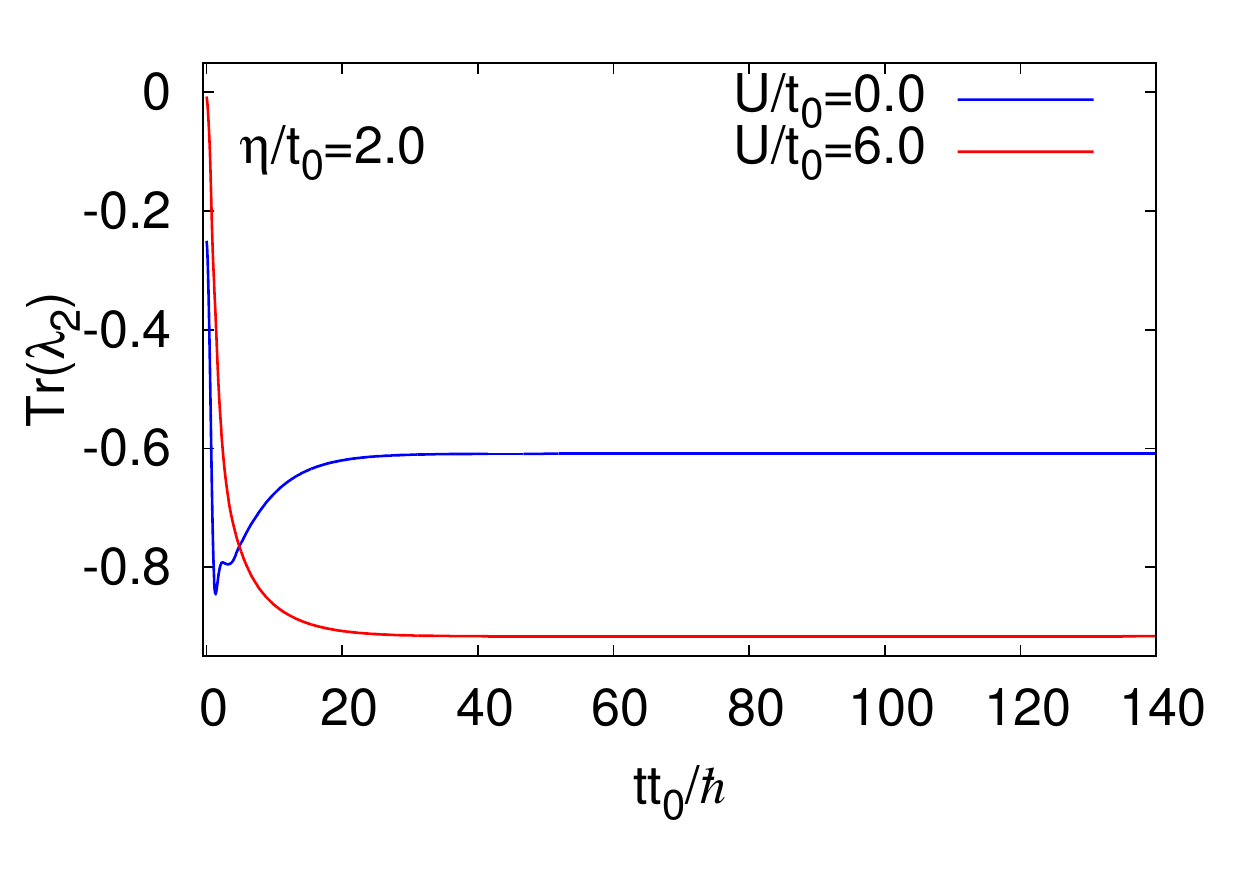}
\caption{\raggedright Two-particle cumulant during the evolution in \fig{dynamics}.}\label{cumulant}
\end{figure}

Does decoherence help us reduce the complexity of the many-body electron problem?  \Fig{cumulant} shows the evolution of the two-particle cumulant ($\text{Tr}[\lambda_2]=\text{Tr}[{}^{(1)}\Gamma^2-{}^{(1)}\Gamma]$, where ${}^{(1)}\Gamma$ is the 1-body electronic density matrix) which measures the importance of 2-body contributions to $\hat{\rho}_e$ that cannot be decomposed in terms of ${}^{(1)}\Gamma$~\cite{Mukherjee}. For an uncorrelated closed electronic system ${}^{(1)}\Gamma^2 = {}^{(1)}\Gamma$ and $\text{Tr}[\lambda_2]=0$. As shown,   instead of reducing the complexity, in this case increasing  $\eta$  (and $U$) enhances the importance of 
 higher order $r$-body electronic density matrices to the BBGKY hierarchy~\cite{bonitz}. 

In conclusion, we have shown that  the  correlation among electrons and the degree of quantum coherence of electronic states are strongly coupled in matter. For arbitrary states, only when the system-bath dynamics is pure dephasing such that \eq{conditions} is satisfied can correlation and decoherence be considered as uncoupled physical phenomena.  Investigating the consequences of this fundamental, ubiquitous, and previously overlooked connection constitutes an emerging challenge in electronic structure and molecular dynamics.

\section*{acknowledgement}
This material is based upon work supported by the National Science Foundation under CHE - 1553939. I.F. thanks Prof.~John Parkhill for helpful discussions.

\section*{Supporting Information Available}
The Supporting Information includes plots for the evolution of the electronic density matrix, the PESs for representative examples, and  a discussion of the mechanisms at play in \fig{dynamics}.

\clearpage

\begin{center}
\textbf{\Large{Understanding the Fundamental Connection Between Electronic Correlation and Decoherence: Supporting Information}}
\end{center}

\setcounter{equation}{0}
\setcounter{figure}{0}
\setcounter{page}{1}
\renewcommand{\theequation}{S\arabic{equation}}
\renewcommand{\thefigure}{S\arabic{figure}}
\renewcommand{\thetable}{S\arabic{table}}
\renewcommand{\thepage}{S\arabic{page}}

\section{Scheme of the adiabatic connection  in Eqs. (1)-(5)} 

\begin{figure}[h]
\centering
\includegraphics[scale=0.7]{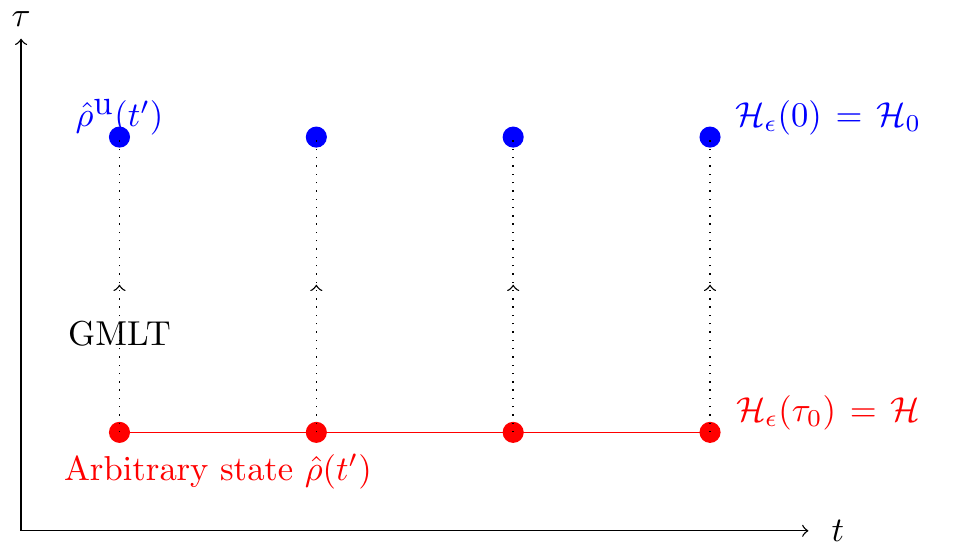}
\caption{\raggedright At each instant of time $t$, the uncorrelated counterpart $\hat{\rho}^\textrm{u}(t)$ of a general electron-nuclear state $\hat{\rho}(t)$ can be obtained by turning off the residual electron-electron interactions $V^S$ adiabatically slow via evolution in the Interaction picture along a fictitious time coordinate $\tau$. }
\label{fig1}
\end{figure}

\section{Decoherence dynamics of the Hubbard-Holstein model}

\noindent
To clarify the molecular mechanisms at play in the decoherence dynamics of the Hubbard-Holstein model shown in Fig.~1, below we discuss the effect of changing $U$ and $\eta$ on the electronic potential energy surfaces (PESs) and on the dynamics of the electronic density matrix.

\subsection{Dynamics of the electronic density matrix  $\hat{\rho}_e(t)$}

\begin{figure}[htbp]
\begin{center}
\includegraphics[width=0.75\textwidth]{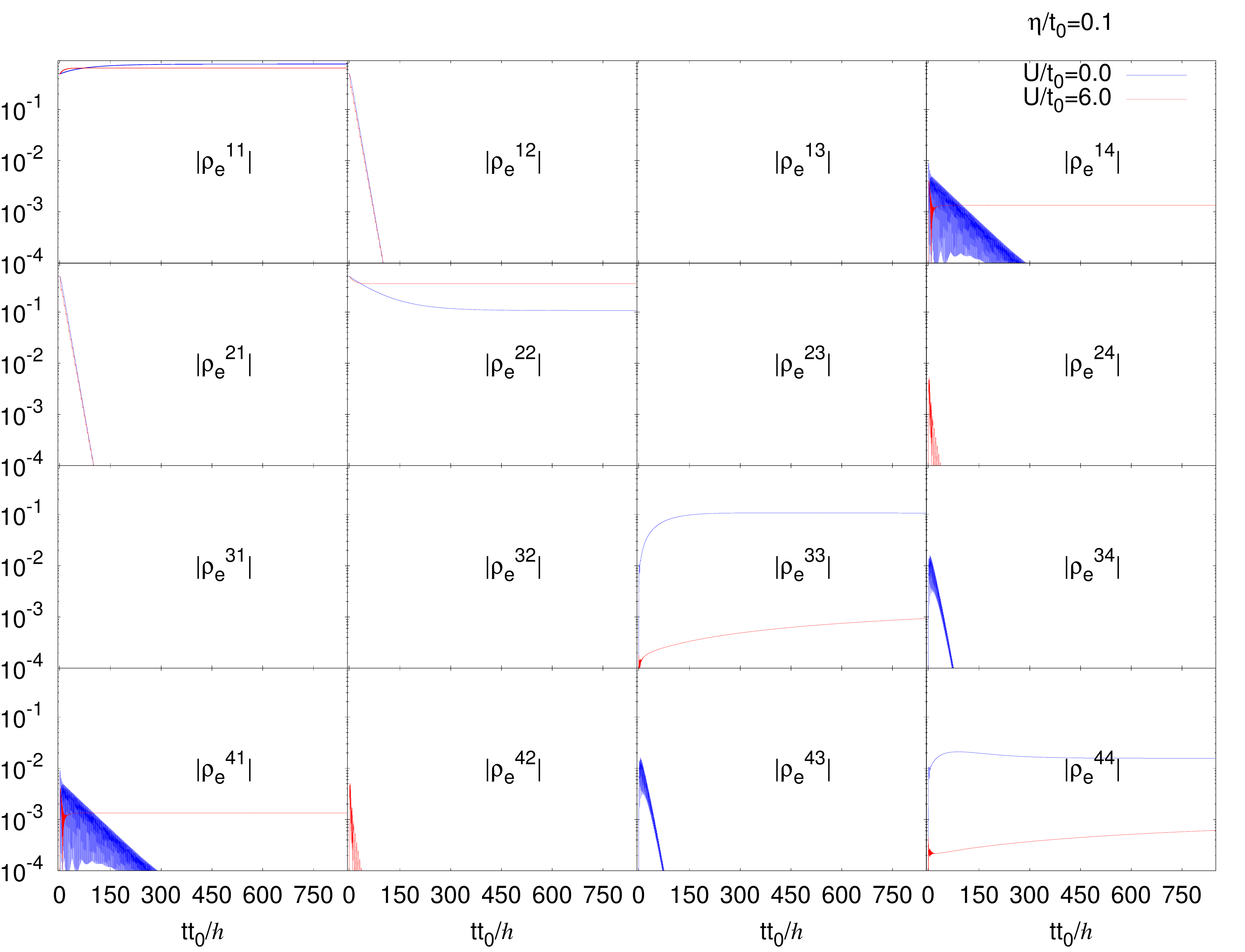}
\includegraphics[width=0.75\textwidth]{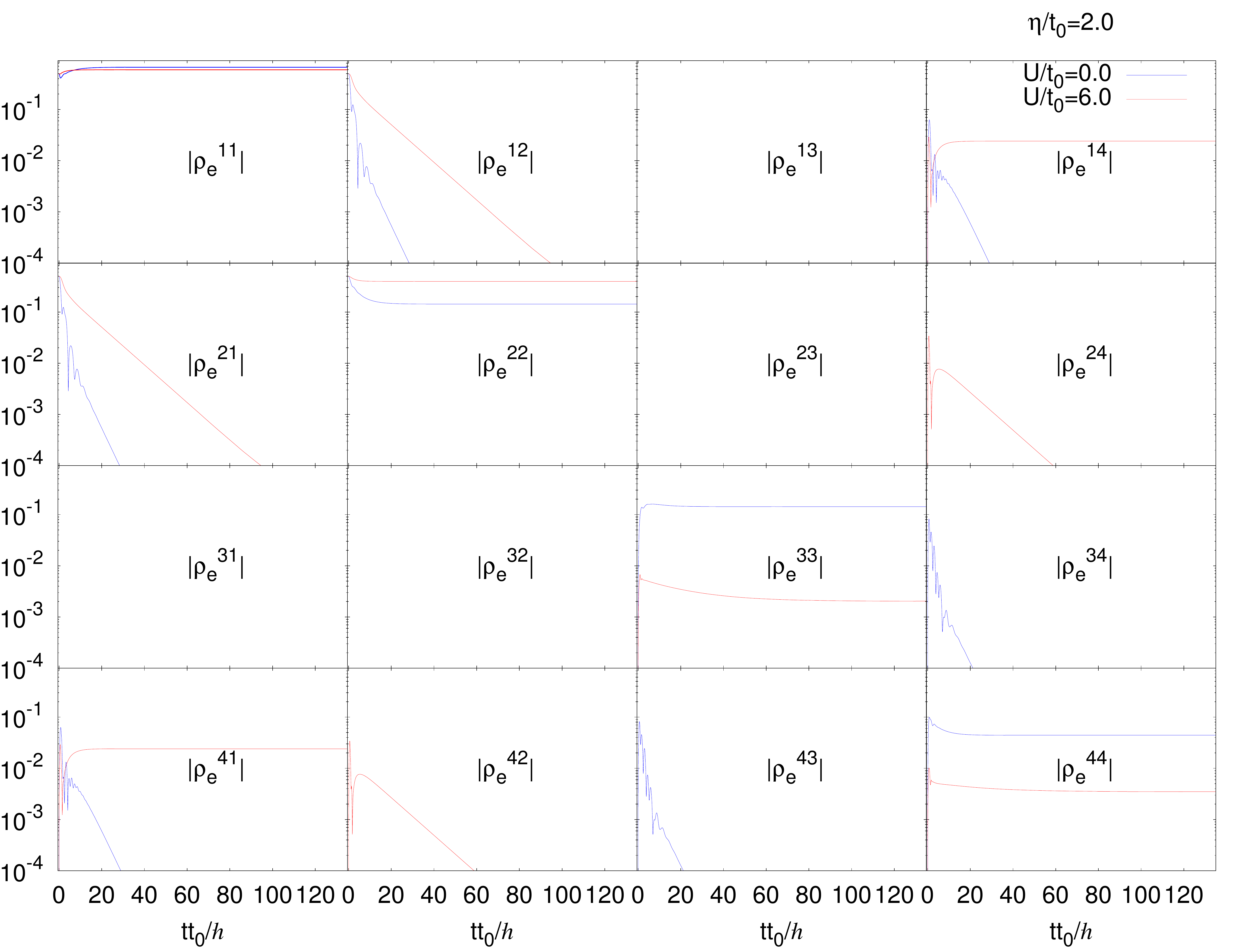}
\end{center}
\caption{\raggedright{\footnotesize{Evolution of the electronic density matrix $\h{\rho}_e(t)$ under the conditions in Fig.~1 with $\eta=0.1 t_0$ (top) and $\eta=2.0 t_0$ (bottom).  Each panel corresponds to a different matrix element of $\h{\rho}_e^{ij}(t)=\langle E_i|\h{\rho}_e(t)|E_j\rangle$  in the eigenbasis of $H^S$ $\{ |E_i\rangle \}$ ordered with increasing energy. Initially $\hat{\rho}_e(0)=\ket{\Psi}\bra{\Psi}$, where $\ket{\Psi}=\frac{1}{\sqrt{2}}(|E_1\ra+|E_2\ra)$.  Timescales for the decay of $\h{\rho}_e^{ij}(t)$ based on an exponential fit are shown in Table~\ref{exponential}. Note that for $\eta=2.0 t_0$  the initial coherence between ground and first excited state ($\h{\rho}_e^{12}(t)$) is protected by increasing $U$. By contrast,  for $\eta=0.1 t_0$ the decay of $\h{\rho}_e^{12}(t)$ is independent of $U$.}}}
\label{coherences}
\end{figure}

\noindent
\Fig{coherences} shows the dynamics  of $\hat{\rho}_e(t)$ in the eigenbasis of $H^S$ for four representative cases ($U=0 t_0,6 t_0$; $\eta=0.1 t_0,2.0 t_0$).  The characteristic relaxation timescale of each matrix element was obtained via an exponential fit and tabulated in Table~\ref{exponential}. As can be seen from \Fig{coherences} and Table~\ref{exponential}, the most important diagonal elements of $\h{\rho}_e(t)$ ($\h{\rho}_e^{11}(t)$ and $\h{\rho}_e^{22}(t)$) decay faster for $U=6 t_0$ than $U=0 t_0$ indicating that increasing the electron-electron interactions generally leads to faster relaxation. For $\eta=0.1 t_0$, the decay of the initial coherence between the ground and first excited state $\h{\rho}_e^{12}(t)$ is largely unaffected by varying $U$. By contrast, for $\eta=2.0 t_0$ the decay of $\h{\rho}_e^{12}(t)$ is slower for $U=6t_0$ than $U=0t_0$ signaling that increasing $U$ protects the coherences between these two electronic energy eigenstates.  Naturally, the thermalization of $\h{\rho}_e(t)$ is  faster for the stronger electron-nuclear coupling ($\eta=2.0 t_0$) than for the weak electron-nuclear coupling ($\eta=0.1 t_0$). As described in the sections below, these features of the dynamics of $\hat{\rho}_e(t)$ can be understood by investigating the effect of changing $U$ and $\eta$ on the PESs, 

Note that the characteristic decay timescales for $\hat{\rho}_e(t)$ in Table~\ref{exponential} are  related to the  decoherence timescales obtained from the purity  (Table~\ref{purityexp}). This is because, the thermalization of the purity
\begin{align}
\label{connection}
P(t)=\text{Tr}[\h{\rho}_e^2(t)]=\sum_{i,j}|\h{\rho}_e^{ij}(t)|^2,
\end{align}
is determined by the individual contributions to the density matrix $\h{\rho}_e^{ij}(t)$. 

\begin{table}[!htb]
    \begin{subtable}{.5\linewidth}
      \centering
        \caption{$\eta/t_0=0.1$}
        \begin{tabular}{|c|c|c|c|c|}
\hline
&\multicolumn{2}{c|}{$U/t_0=0$} & \multicolumn{2}{c|}{$U/t_0=6$} \\
\hline 
& $p$ & $\tau(t_0/\hbar)$ & $p$ & $\tau (t_0/\hbar)$ \\
\hline
$\h{\rho}_e^{11}$ & {\color{blue}-0.2763} & {\color{blue}92.42} & {\color{red}-0.1568} & {\color{red}11.65}\\
$\h{\rho}_e^{12}$ & {\color{blue}0.5449} & {\color{blue}12.82} & {\color{red}0.4993} & {\color{red}10.91}\\
$\h{\rho}_e^{13}$ & {\color{blue}0} & {\color{blue}} & {\color{red}0} & {\color{red}}\\
$\h{\rho}_e^{14}$ & {\color{blue}0.0038} & {\color{blue}67.34} & {\color{red}0} & \\
$\h{\rho}_e^{22}$ & {\color{blue}0.3861} & {\color{blue}80.12} & {\color{red}0.1571} & {\color{red}12.47}\\
$\h{\rho}_e^{23}$ & {\color{blue}0} & {\color{blue}} & {\color{red}0} & {\color{red}}\\
$\h{\rho}_e^{24}$ & {\color{blue}0} & {\color{blue}} & {\color{red}0.0036} & {\color{red}7.923}\\
$\h{\rho}_e^{33}$ & {\color{blue}-0.1055} & {\color{blue}63.45} & {\color{red}-0.0018} & {\color{red}1533}\\
$\h{\rho}_e^{34}$ & {\color{blue}0.0116} & {\color{blue}22.24} & {\color{red}0} & {\color{red}}\\
$\h{\rho}_e^{44}$ & {\color{blue}0.0032} & {\color{blue}319.5} & {\color{red}-0.0009} & {\color{red}1550}\\
\hline
        \end{tabular}
    \end{subtable}%
    \begin{subtable}{.5\linewidth}
      \centering
        \caption{$\eta/t_0=2.0$}
        \begin{tabular}{|c|c|c|c|c|}
\hline
&\multicolumn{2}{c|}{$U/t_0=0$} & \multicolumn{2}{c|}{$U/t_0=6$} \\
\hline 
& $p$ & $\tau (t_0/\hbar)$ & $p$ & $\tau (t_0/\hbar)$ \\
\hline
$\h{\rho}_e^{11}$ & {\color{blue}-0.2633} & {\color{blue}6.134} & {\color{red}-0.1162} & {\color{red}3.025}\\
$\h{\rho}_e^{12}$ & {\color{blue}0.5289} & {\color{blue}1.244} & {\color{red}0.4515} & {\color{red}7.679}\\
$\h{\rho}_e^{13}$ & {\color{blue}0} & {\color{blue}} & {\color{red}0} & {\color{red}}\\
$\h{\rho}_e^{14}$ & {\color{blue}0.0372} & {\color{blue}2.810} & {\color{red}-0.0166} & {\color{red}5.880}\\
$\h{\rho}_e^{22}$ & {\color{blue}0.3265} & {\color{blue}3.723} & {\color{red}0.1146} & {\color{red}2.616}\\
$\h{\rho}_e^{23}$ & {\color{blue}0} & {\color{blue}} & {\color{red}0} & {\color{red}}\\
$\h{\rho}_e^{24}$ & {\color{blue}0} & {\color{blue}} & {\color{red}0.0112} & {\color{red}13.75}\\
$\h{\rho}_e^{33}$ & {\color{blue}0.0009} & {\color{blue}123.8} & {\color{red}0.0033} & {\color{red}23.17}\\
$\h{\rho}_e^{34}$ & {\color{blue}0.0565} & {\color{blue}2.601} & {\color{red}0} & {\color{red}}\\
$\h{\rho}_e^{44}$ & {\color{blue}0.0343} & {\color{blue}7.220} & {\color{red}0.0023} & {\color{red}19.02}\\
\hline
        \end{tabular}
    \end{subtable} 
\caption{\raggedright Characteristic timescales in the thermalization of the density matrix of the Hubbard-Holstein model  shown in \fig{coherences}. The data corresponds to an exponential fit of the decay  $\h{\rho}^{ij}_e(t)-\h{\rho}_e^{ij}(\infty)=p e^{-\frac{t}{\tau}}$, where $\tau$ is the timescale and $p$ its weight.}
\label{exponential}
\end{table}

\subsection{Effect of changing $U$ and $\eta$ on the PESs}

\begin{figure}[htb!]
\begin{center}
\includegraphics[width=0.3\linewidth]{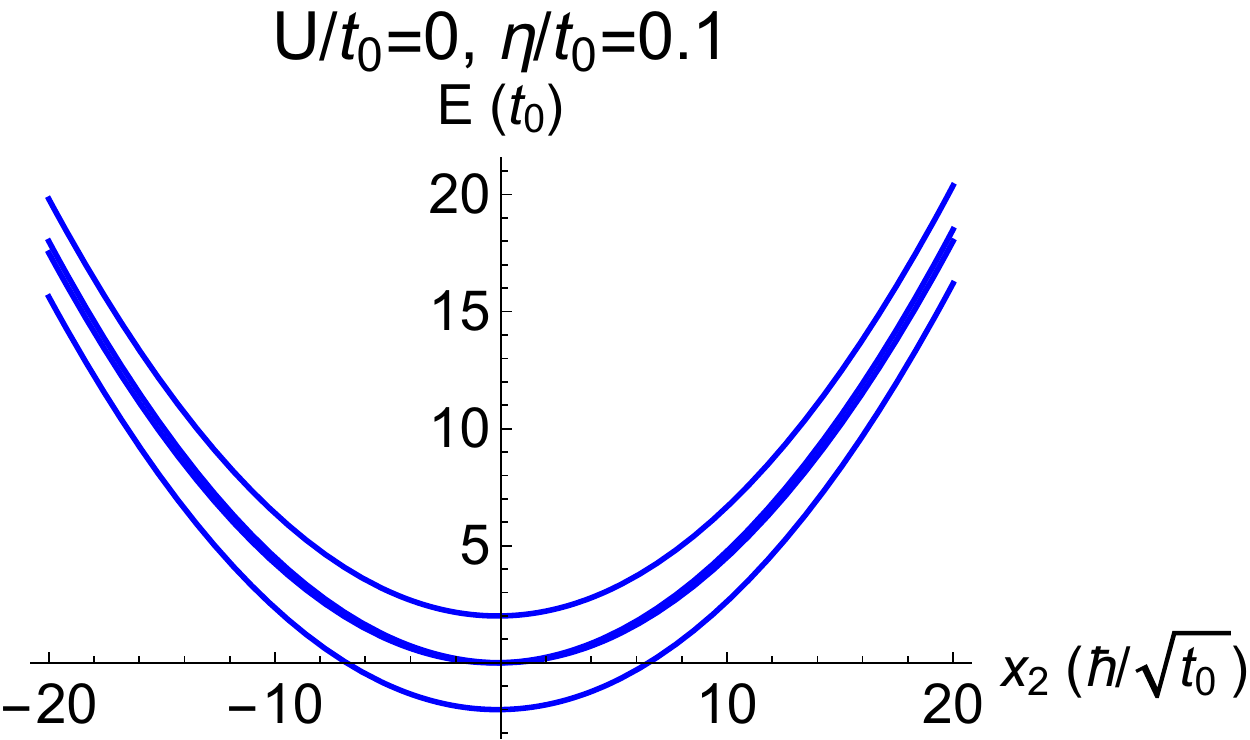}
\includegraphics[width=0.3\linewidth]{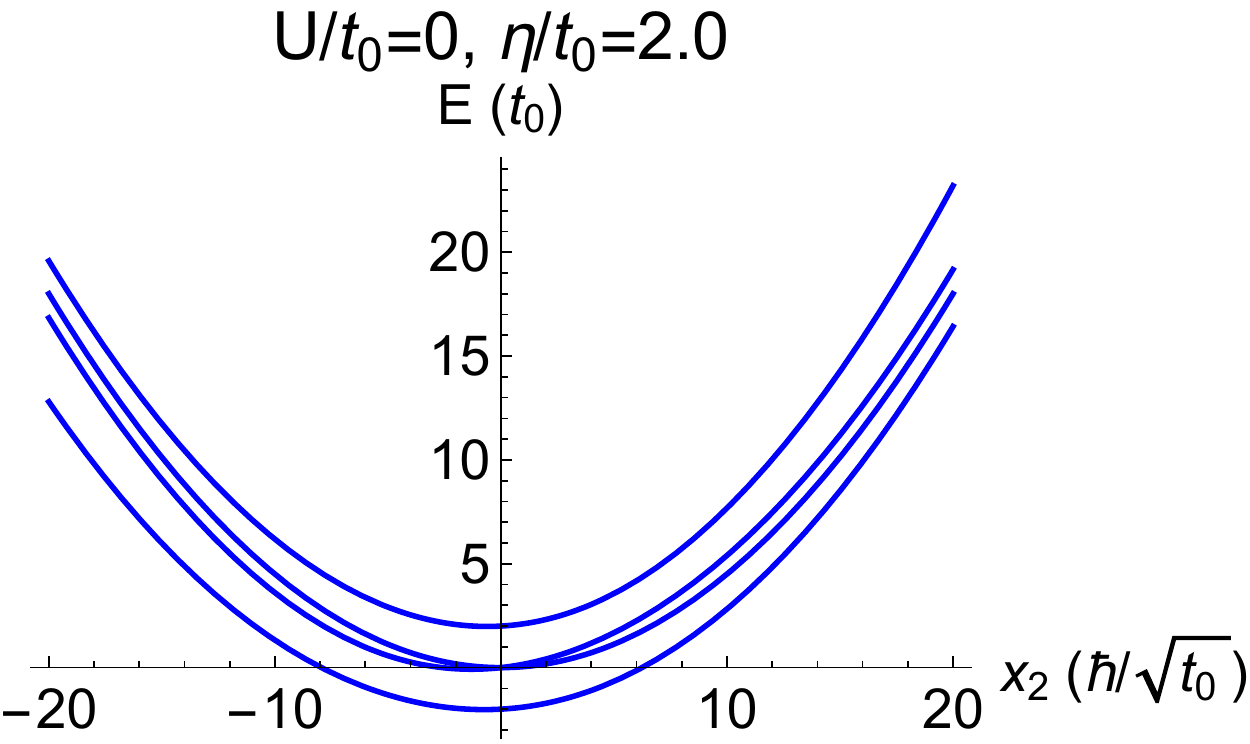}

\includegraphics[width=0.3\linewidth]{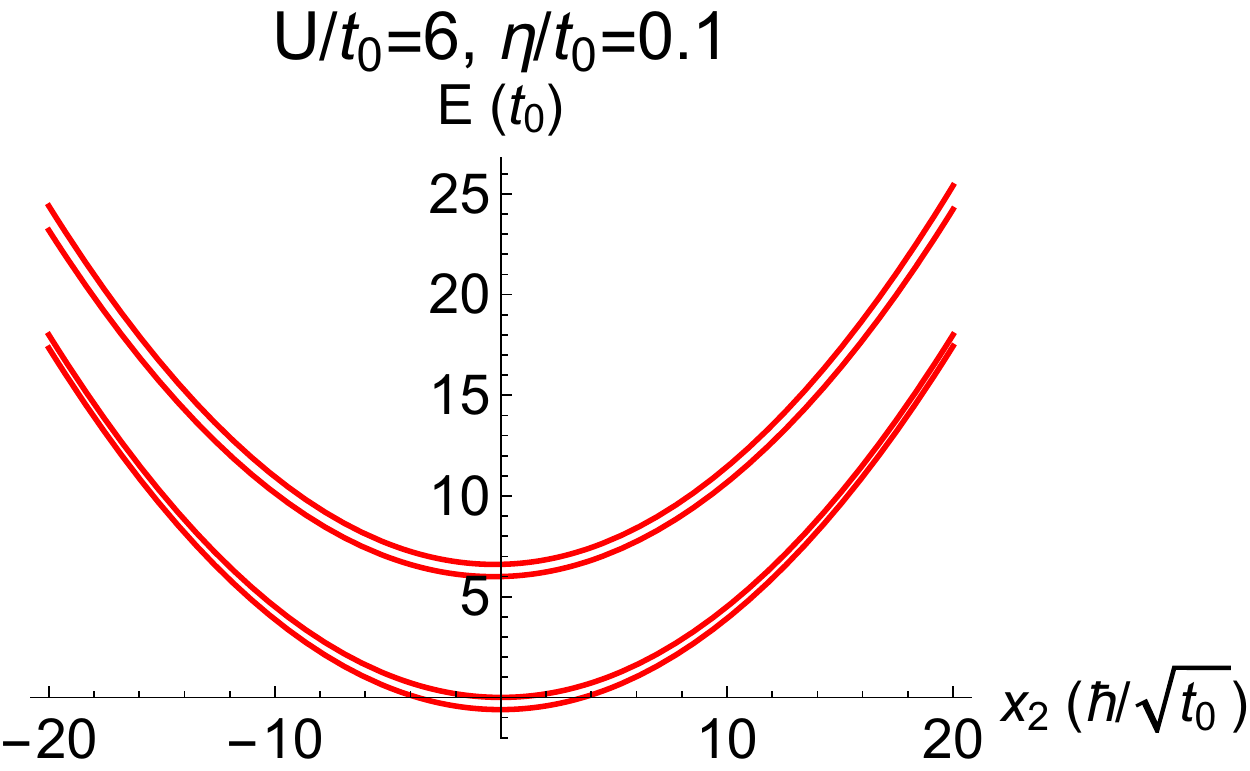}
\includegraphics[width=0.3\linewidth]{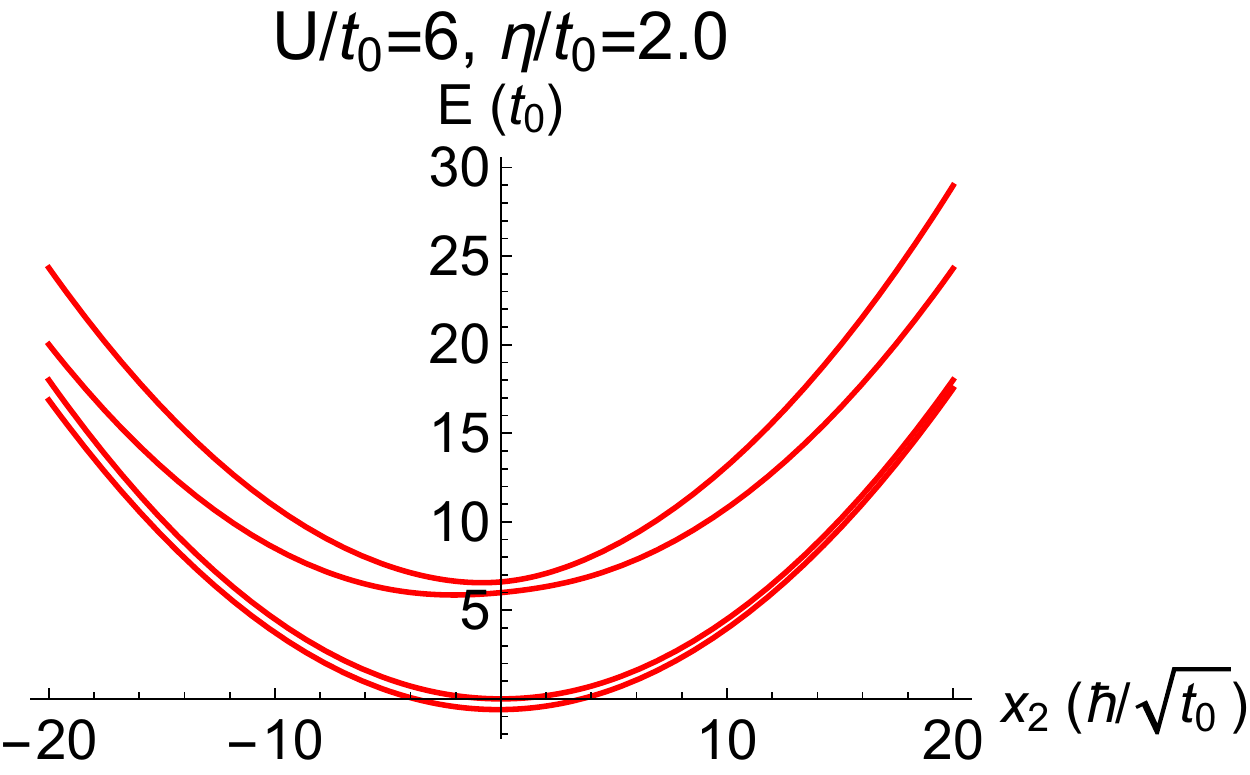}

\includegraphics[width=0.3\linewidth]{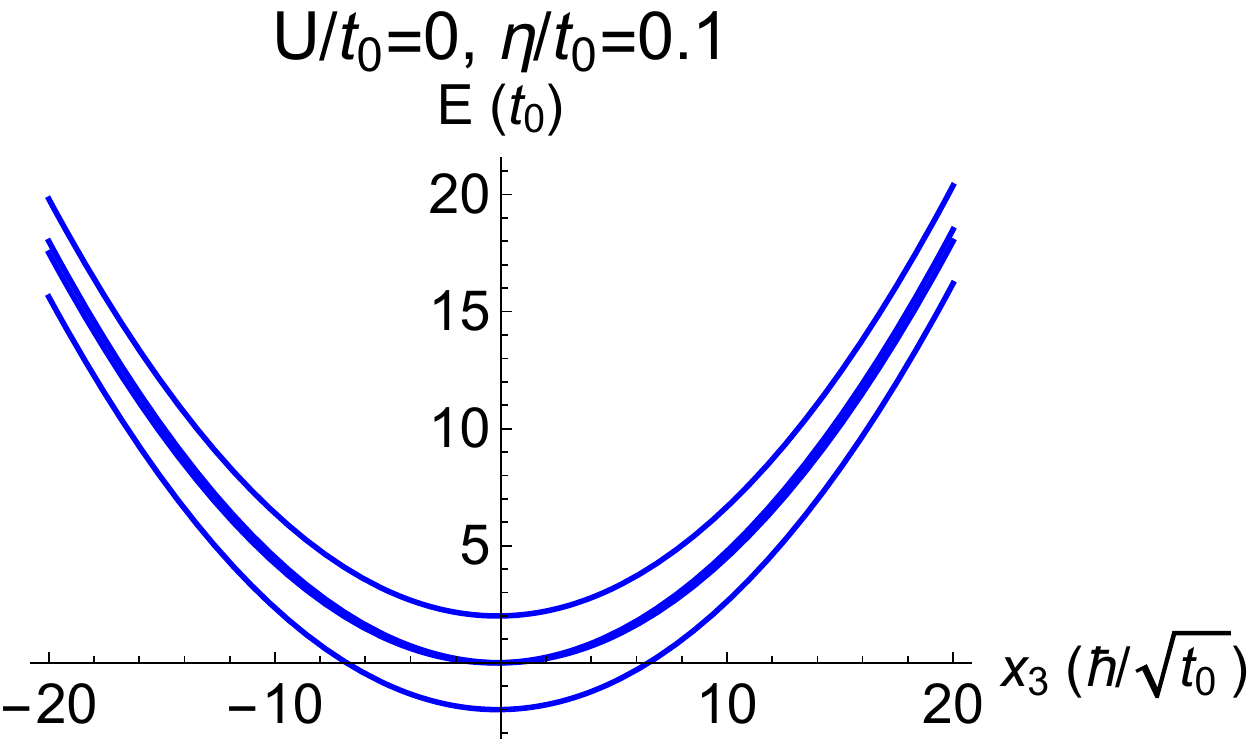}
\includegraphics[width=0.3\linewidth]{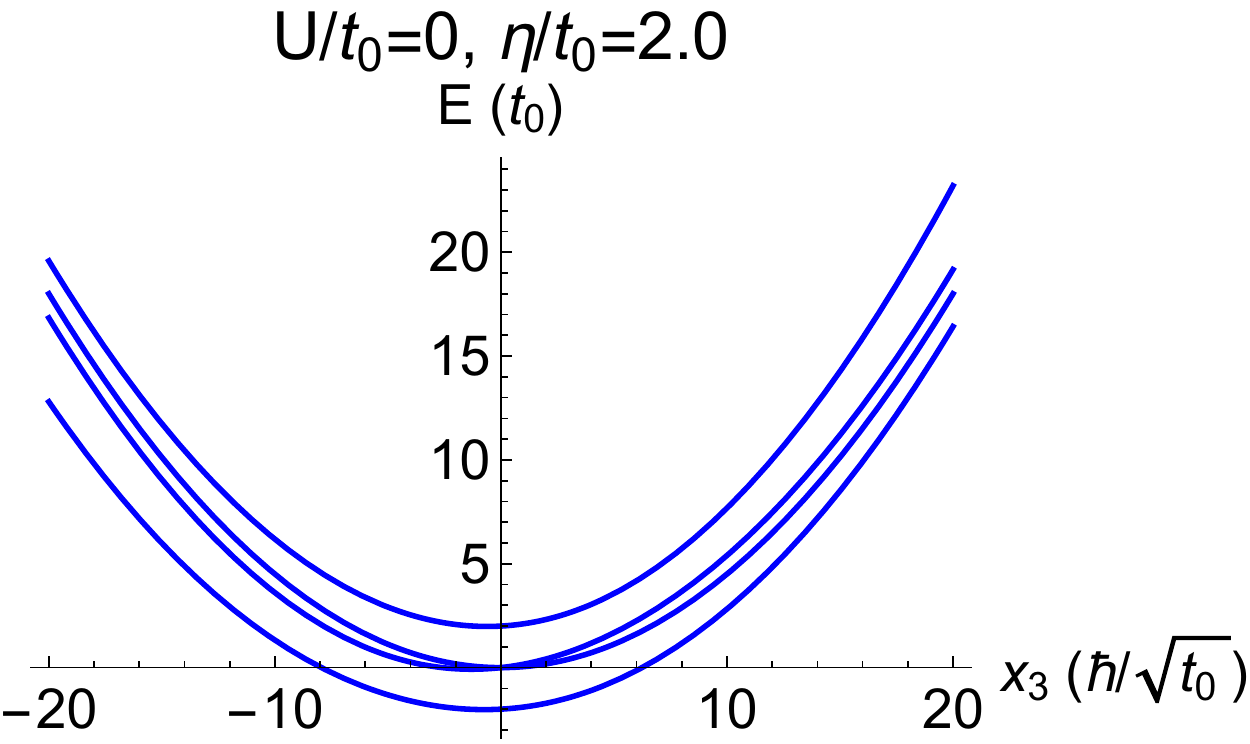}

\includegraphics[width=0.3\linewidth]{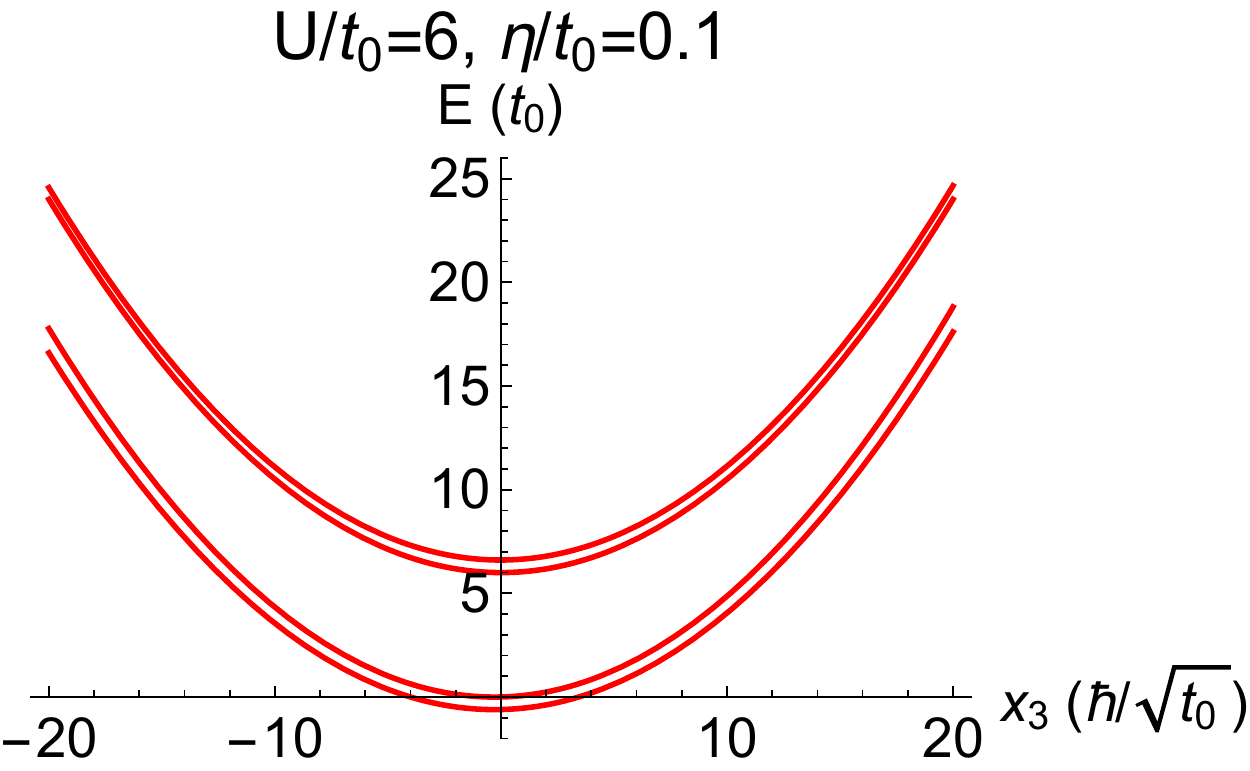}
\includegraphics[width=0.3\linewidth]{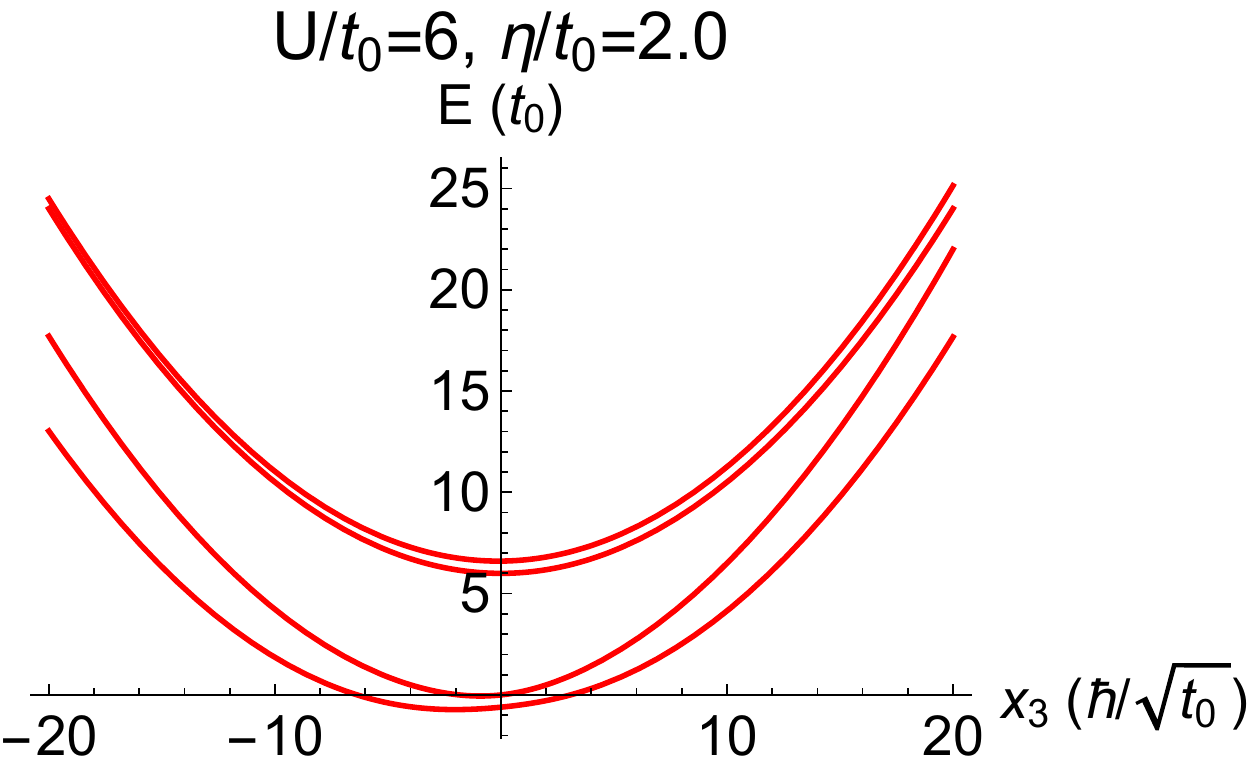}
\caption{\raggedright Projections of the PESs for the simplified Hubbard-Holstein model defined in \eq{eq:modifiedHH} along $x_{2}$ (with  $x_{1}=x_{3}=x_{4}=0$), $E^{(n)}(x_2)$,  and $x_{3}$ (with  $x_{1}=x_{2}=x_{4}=0$),  $E^{(n)}(x_3)$, with $n=1,\cdots, 4$. The PESs along $x_4$ (or $x_1$) are identical to those along  $x_3$ (or $x_2$). Note that  with an increase in $U$ the ground and first excited (or second and third excited) states come closer in energy. }\label{PES}
\end{center}
\end{figure}

\noindent
Consider now the effect of changing $U$ and $\eta$ on the PESs of the Hubbard-Holstein model. For definitiveness, we focus on the limiting case where the molecule is coupled to just one of the harmonic oscillators in each independent bath. In this case, the Hubbard-Holstein model  Eqs.~(14)-(16) reduces to:
\begin{align}
\begin{split}
\mc{H} = & -t_0\sum_{\sigma\in\{\uparrow,\downarrow\}}(\h{d}^{\dagger}_{1\sigma}\h{d}_{2\sigma}+\h{d}^{\dagger}_{2\sigma}\h{d}_{1\sigma})
+U(\h{n}_{1\uparrow}\h{n}_{1\downarrow}+\h{n}_{2\uparrow}\h{n}_{2\downarrow}) + 
\sum_{m=1}^4\left(\frac{p_{m}^2}{2}+\frac{1}{2}\omega^2x^2_{m}\right)\\
& +c_{1}x_{1}\h{n}_{1\uparrow}\h{n}_{1\downarrow}+c_{2}x_{2}\h{n}_{2\uparrow}\h{n}_{2\downarrow}+c_{3}x_{3}\h{n}_{1\uparrow}\h{n}_{2\downarrow}+c_{4}x_{4}\h{n}_{1\downarrow}\h{n}_{2\uparrow},
\label{eq:modifiedHH}
\end{split}
\end{align}
where the frequency is taken to be at the  peak of the spectral density ($\hbar \omega=0.3 t_0$). The strength of the electron-nuclear coupling is determined by $\{c_{m}\}$, chosen to be $c_{m}=\omega\sqrt{\frac{\eta}{\pi}}$. While this system  does not correspond exactly to the system that is modeled through the HEOM approach, it does allow extracting qualitative understanding of the effect of changing $U$ and $\eta$ on the PESs.  \Fig{PES} shows one-dimensional projections of the four PESs for this model along $x_{2}$ (with  $x_{1}=x_{3}=x_{4}=0$) and $x_{3}$ (with  $x_{1}=x_{2}=x_{4}=0$) for different choices of $U$ and $\eta$. Note that for this simplified model the PESs  along $x_{3}$ and $x_{4}$ (or $x_{1}$ and $x_{2}$) coincide. As can be seen in \fig{PES}, for weak electron-nuclear couplings ($\eta=0.1 t_0$) the 4 PESs are very similar, while for the stronger $\eta=2.0 t_0$ the minimum and curvature of the PESs generally differ. In both cases, the effect of increasing $U$ is to bring the ground and first excited state (or second and third excited states) closer together in energy, and to reduce the difference in curvature of the PESs associated with the ground and first excited state.

\begin{figure}[htbp]
\begin{center}
\includegraphics[width=0.4\linewidth]{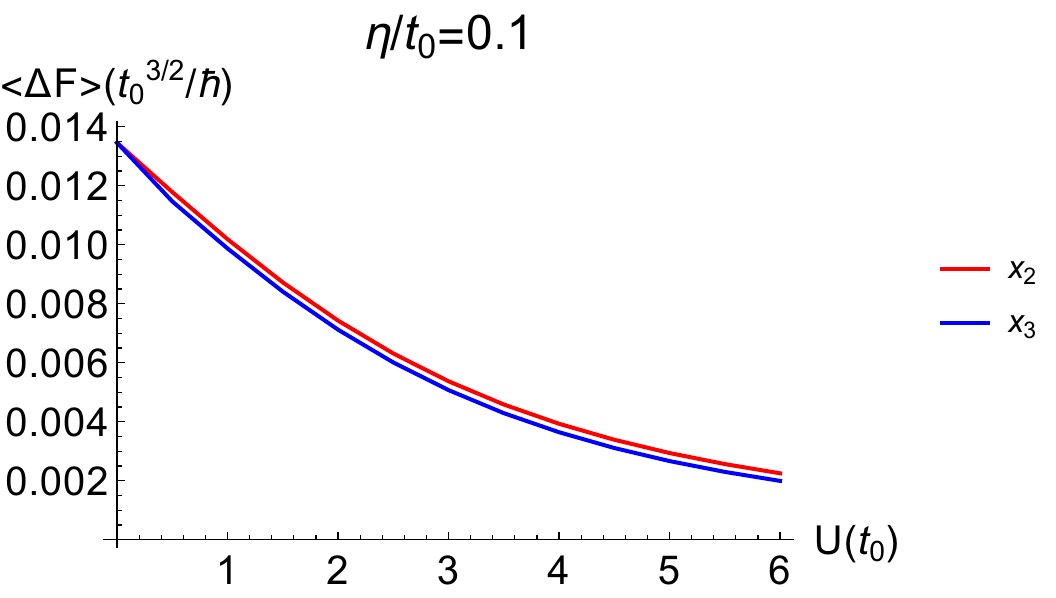}
\includegraphics[width=0.4\linewidth]{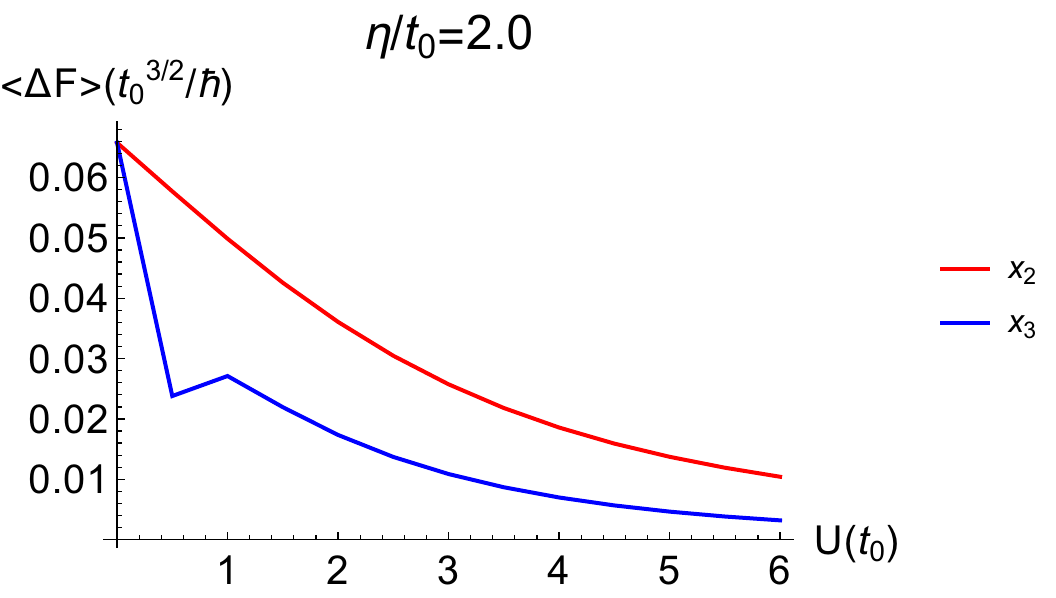}
\caption{\raggedright Average of the difference in forces $\la\Delta F\ra$ (\eq{average})  between the ground and first excited state for different $U$ and $\eta$ along two different nuclear coordinates $x_{2}$ (with  $x_{1}=x_{3}=x_{4}=0$) and $x_{3}$ (with  $x_{1}=x_{2}=x_{4}=0$). The decrease in  $\la\Delta F\ra$ with an increase in $U$ can increase the lifetime of coherences between these two states.}\label{forces}
\end{center}
\end{figure}

The reduction in curvature difference can be quantified through
\begin{align}
\label{average}
\la \Delta F\ra=\la F_1-F_2\ra=\frac{\dsp{\sum_n} e^{-\beta \epsilon_n} \la \chi_n|(F_1-F_2)|\chi_n\ra}{\dsp{\sum_n} e^{-\beta \epsilon_n}},
\end{align}
which measures the average of the difference in the curvatures $F_n=d E^{(n)}(x_i)/d x_i$  between ground ($n=1$) and first ($n=2$) excited state along a particular nuclear coordinate $x_i$, where $E^{(n)}(x_i)$ is the adiabatic PES of the $n$-th electronic state along $x_i$. The average is taken over the initial nuclear thermal state, where $\epsilon_n$ and $|\chi_n\ra$ are the eigenvalues and eigenfunctions of the $n$-th  harmonic oscillator level, and $\beta$ the inverse temperature. As shown in \fig{forces},  increasing $U$ generally leads to a decrease in the difference in curvature between the ground and first excited state along all nuclear directions. Note that the $\la \Delta F\ra$ for $\eta=0.1 t_0$ are $\sim 5$ times smaller than those for $\eta=2.0 t_0$.

\begin{figure}[htbp]
\begin{center}
\includegraphics[width=0.4\linewidth]{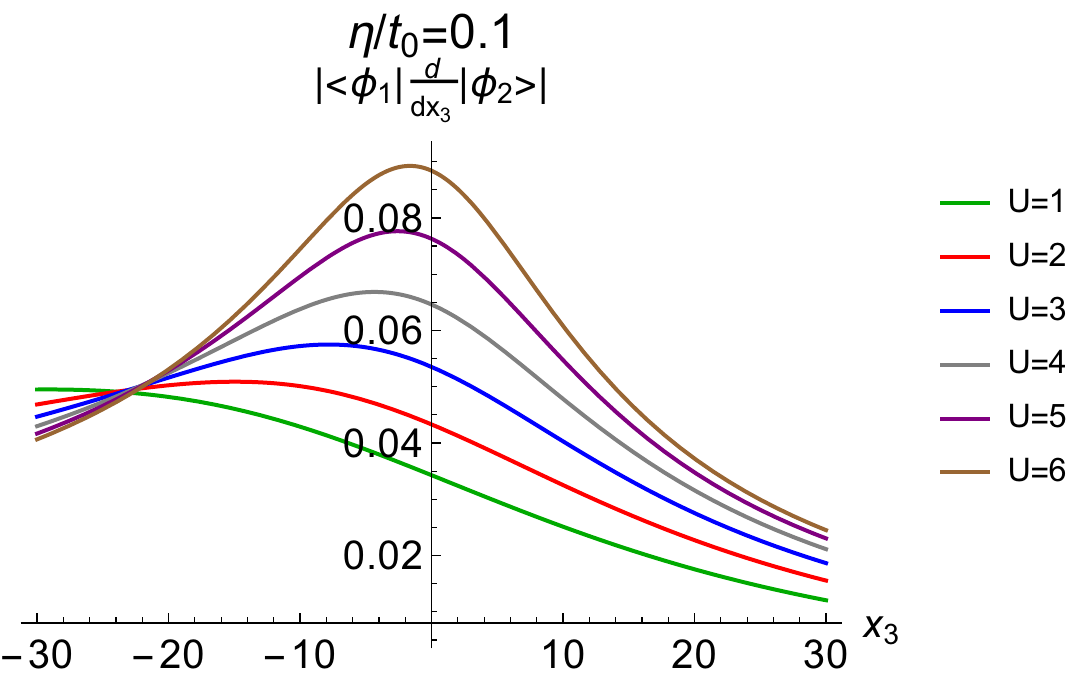}
\includegraphics[width=0.4\linewidth]{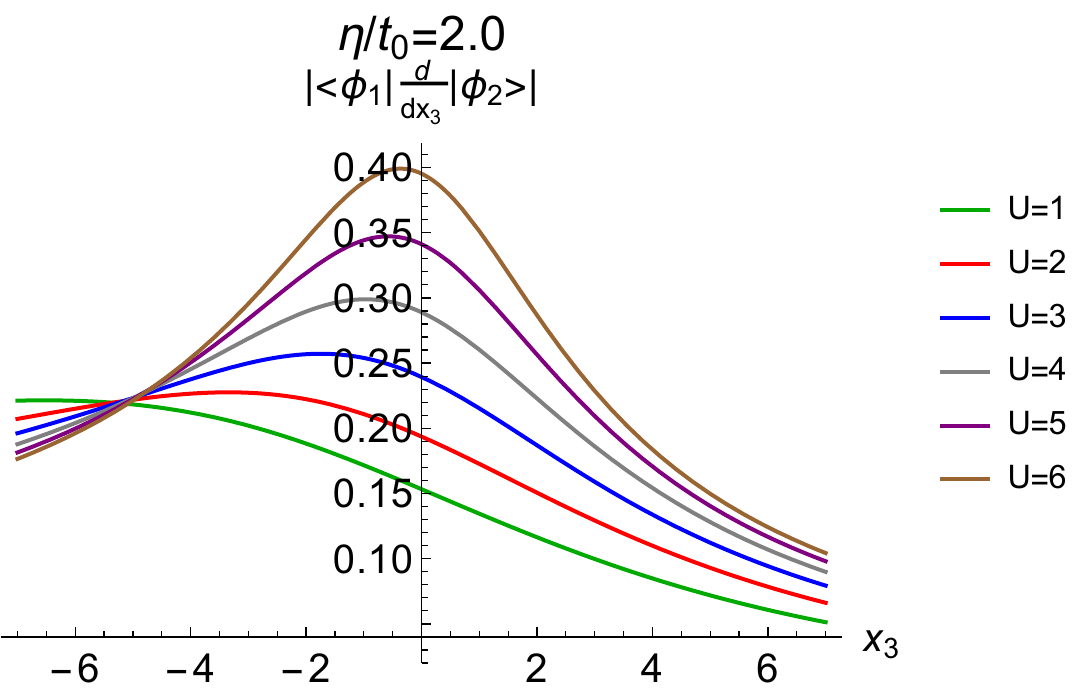}
\caption{\raggedright Nonadiabatic couplings (NACs) [\eq{naceq}] between the ground ($|\phi_1\ra$) and first excited BO electronic state ($|\phi_2\ra$) as a function of the nuclear coordinate $x_{3}$ (or $x_4$) for $\eta=0.1 t_0$ (left) and $\eta=2.0 t_0$ (right). The NACs vanish along the nuclear coordinate $x_{1}$ and $x_2$. As $U$ increases, $|\phi_1\ra$ and $|\phi_2\ra$ come closer in energy (see \fig{PES}) causing an increase in the NACs. }\label{NAC}
\end{center}
\end{figure}

The decrease in energy difference between the ground and first excited state with increasing $U$, causes the nonadiabatic couplings (NACs), $d_{12}$, between these two states to increase (see \fig{NAC}).   The NACs between electronic eigenstates  $|\phi_1(x_m)\ra$  and $|\phi_2(x_m)\ra$  are defined by~\cite{Tully}
\begin{align}
\label{naceq}
d_{12}= |\la \phi_1(x_m)|\frac{\partial}{\partial x_m}|\phi_2(x_m)\ra|.
\end{align}
Here $|\phi_i(x_m)\ra$ refers to the $i$-th Born-Oppenheimer (BO) electronic eigenstate of the Hamiltonian (obtained by diagonalizing everything in \eq{eq:modifiedHH} except the nuclear kinetic energy). The $d_{12}$  measure the coupling between two electronic levels via nuclear motion. An increase in the NACs leads to increased excitation of the electrons via nuclear dynamics. As shown in \fig{NAC}, the NACs are $\sim 5$ times larger for the case of stronger electron-nuclear couplings. Further, the NACs increase significantly with an increase in $U$ for both values of $\eta$ considered  as a result of the energy levels coming closer together. 


\subsection{Two competing decoherence mechanisms}

As detailed above, increasing $U$ brings the ground and first  excited state  closer together in energy, and  reduces the difference in curvature between their PESs. As we now discuss, these two effects on the PESs lead to competing decoherence mechanisms that underlie the dynamics in Fig.~1. 


\subsubsection{Increasing $U$ decreases the rate of decoherence because it reduces the difference in curvature between the PESs}
In pure electron-nuclear systems, decoherence arises because of nuclear evolution in alternative PESs. To see this, consider the electronic density matrix associated with a general entangled vibronic state $\ket{\Omega (t)} = \sum_n \ket{E_n} \ket{\chi_n(t)}$, 
\begin{equation}
\label{eq:rrho}
\hat{\rho}_e(t) = \textrm{Tr}_B\{\ket{\Omega}\bra{\Omega}\}
=\sum_{nm} \langle \chi_m (t) | \chi_n (t) \rangle  | E_n\rangle \langle E_m |,
\end{equation}
where the trace is over the environmental degrees of freedom, the $\{\ket{E_n}\}$ are the  eigenstates of $H^{S}$ and the  $\ket{\chi_n(t)}$ is the nuclear wavepacket associated with the $n$-th electronic state. Note that the coherences between electronic eigenstates (the off-diagonal elements in $\hat{\rho}_e(t)$) are determined by the  nuclear overlaps $S_{nm}(t)=\langle  \chi_m (t)| \chi_n (t) \rangle $. Thus, the loss of coherences in $\hat{\rho}_e(t)$ can be interpreted as the result of the decay of the $S_{nm}$ during the coupled electron-nuclear evolution~\cite{Prezdho_QuantumTransitionRates,Franco_Decoherencedynamics}.  Anything that leads to a decay in the nuclear overlaps (anharmonicities in the PES, nuclear motion in high-dimensional space, etc.) leads to decoherence.  Standard measures of decoherence capture precisely this. For example, the purity, the measure of decoherence that we focus on here, is given by 
\begin{equation}
\label{eq:purity}
P(t) =   \sum_{nm} |\langle \chi_m(t)|\chi_n(t)\rangle |^2
\end{equation}
and decays with the overlaps between the environmental states $S_{nm}$.

The effect of increasing $U$ is to reduce the difference in the curvature between the ground and first excited state  as revealed by  $\langle \Delta F \rangle$ in \fig{forces}. This effect is particularly important for $\eta=2.0t_0$  while for $\eta =0.1 t_0$ the four PESs are very similar to one another even for $U=0 t_0$.  Given the initial coherence between the ground and first excited state, this reduction in the curvature is expected to slow down the decoherence because it leads to a slower decay in the overlap of the nuclear wavepackets $S_{12}$ associated with these two states for each member of the initial thermal ensemble.  For  $\eta=2.0 t_0$, this feature is clearly reflected in the dynamics of $\h{\rho}_e^{12}(t)$ that shows an increase in coherence time from 1.2 to 7.7 $t_0/\hbar$ as  $U$ changes from $0 t_0$ to $6 t_0$ (\fig{coherences}). By contrast,  for $\eta =0.1 t_0$, the shape of the PESs is mostly unaltered by varying $U$, hence the reason why in \fig{coherences} the $\h{\rho}_e^{12}(t)$ decay at approximately the same rate for $U=0 t_0$ and $U=6 t_0$.

\subsubsection{Increasing $U$ increases the rate of decoherence because it reduces the energy difference between  electronic states}
By reducing the energy difference between the ground and first excited state, increasing $U$ introduces an additional decoherence mechanism in the dynamics that arises because the nuclei are initially prepared in a thermal incoherent state. Specifically, as the energy difference between levels is reduced with increasing $U$,  the NACs between such levels  increase (see \fig{NAC}). This increase in the coupling leads to an enhanced excitation of the electronic degrees of freedom by the nuclear dynamics.  Now, excitation of a coherent system by an incoherent bath leads to decoherence~\cite{vision}.  Therefore, the enhanced excitation of the electronic subsystem by the thermal incoherent nuclear state leads to an increased rate of decoherence. For small $U$, the PESs in \fig{PES} are well separated in energy and this mechanism is suppressed. As $U$ increases this mechanism becomes increasingly important leading to faster decoherence.
 
 Note that for $\eta =0.1 t_0$ this mechanism is expected to be dominant since the PESs in this case are essentially parallel. This explains the comparatively long decoherence time observed when $U=0 t_0$. By contrast, for  $\eta =2.0 t_0$ both decoherence mechanisms are expected to play a role. This explains why the decoherence is significantly faster for $\eta =2.0 t_0$  with respect to $\eta =0.1 t_0$ for all $U$ considered.

\section{Timescales in the purity dynamics}

\begin{figure}[htbp]
\begin{center}
\includegraphics[width=0.40\linewidth]{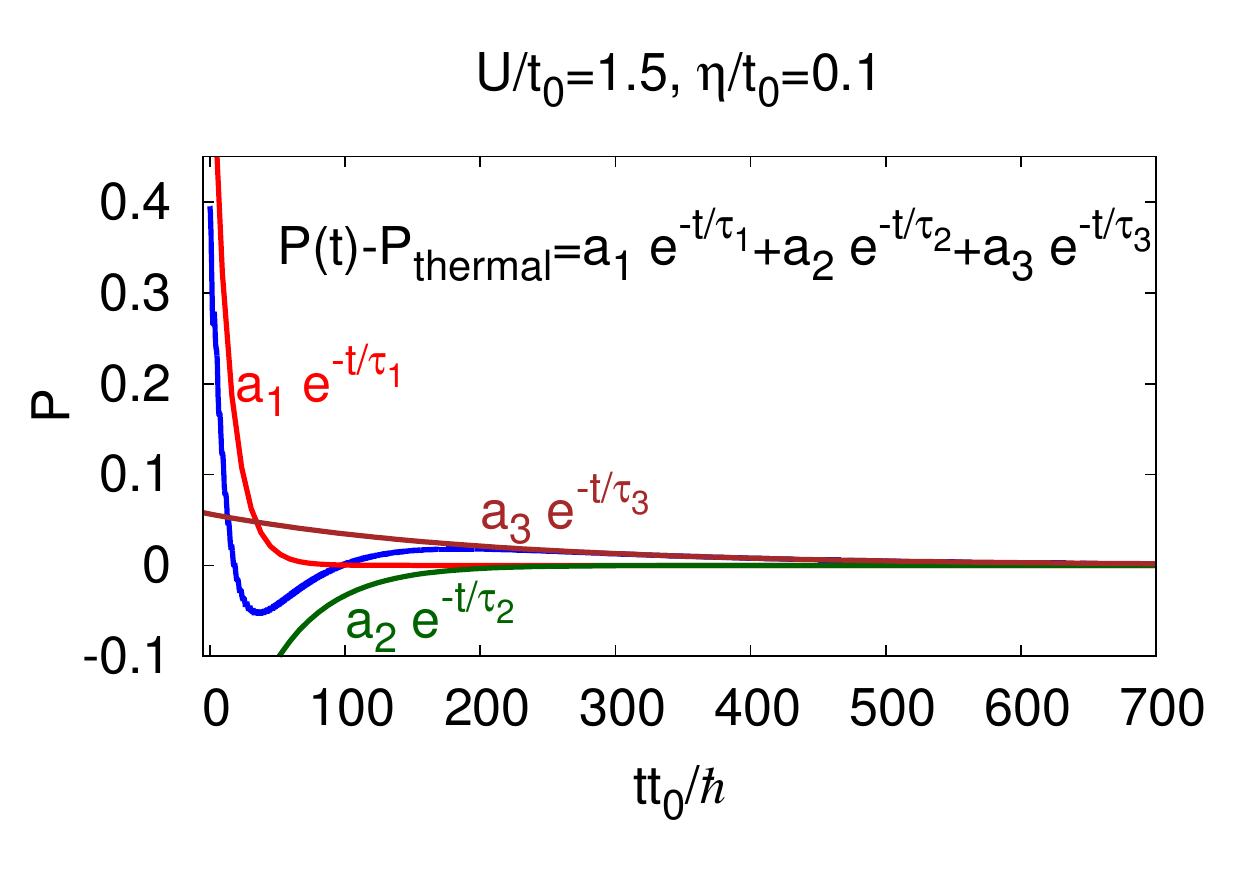}
\includegraphics[width=0.40\linewidth]{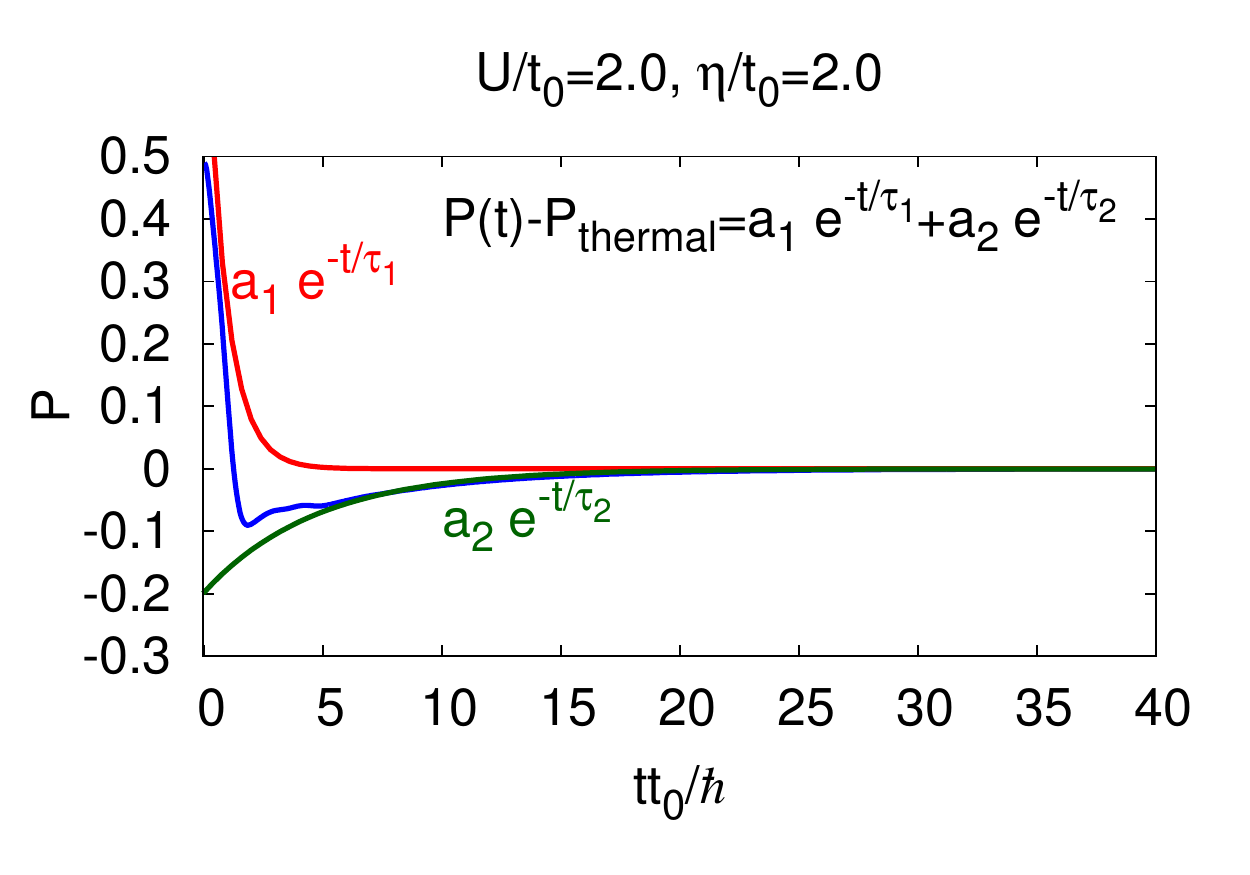}
\caption{\raggedright Pictorial representation of the timescales associated with the purity dynamics for $\eta=0.1 t_0$ (left) and $\eta=2.0 t_0$ (right) is exemplified in a particular case. For $\eta=0.1 t_0$, the three timescales $\tau_1,\tau_2,\tau_3$ can be associated with the initial decay, growth and final thermalization to thermal purity. For $\eta=2.0 t_0$, $\tau_1$ captures the initial decay followed by a growth/decay captured by timescale $\tau_2$.}\label{fits}
\end{center}
\end{figure}

\noindent
Here we illustrate the meaning of the characteristic decoherence timescales $\tau_1$ and  $\tau_2$ in Fig.~1 through a particular example.  \Fig{fits} shows the three timescales $\tau_1, \tau_2$ and $\tau_3$ associated with the purity dynamics obtained through a tri-exponential fit. The initial decay is captured by $\tau_1$ which has a significant contribution to the purity (see Table~\ref{purityexp}), followed by a growth in purity captured by $\tau_2$ and subsequently a final decay captured by $\tau_3$.  The contribution of the $\tau_3$ timescale is negligible in the $\eta=2.0t_0$ dynamics and quite small for $\eta=0.1t_0$ (see Table~\ref{purityexp}), and is not included in Fig.~1. 

\begin{table}[htbp]
\caption{\raggedright Characteristic timescales $\tau_i$ in the dynamics of the purity in Fig.~1 extracted from a triexponential fit $P(t)-P_{\text{thermal}}=\dsp{\sum_{i=1}^3} a_i \exp(-t/\tau_i)$, where $a_i$ quantifies the weight of each contribution. Note that $a_3=0$ for $\eta=2.0t_0$ and small for $\eta=0.1t_0$.}
\centering
\begin{subtable}{.45\textwidth}
\caption{$\eta/t_0=0.1$}
\begin{tabular}{|c|c|c|c|c|c|c|}
\hline
&\multicolumn{6}{c|}{$P(t)-P_{\text{thermal}}=\dsp{\sum_{i=1}^3} a_i \exp(-t/\tau_i)$}\\
\hline 
$U$& $a_1$ & $\tau_1 (t_0/\hbar)$ & $a_2$ & $\tau_2 (t_0/\hbar)$ & $a_3$ & $\tau_3 (t_0/\hbar)$ \\
\hline
$0.0 t_0$ & 0.631 &   9.30 & -0.232 & 115.71& 0 & \\
$0.5 t_0$ & 0.631 & 10.45 & -0.231 &  71.38 & 0 & \\
$1.0 t_0$ & 0.630 & 11.74 & -0.310 &  61.24 & 0.076 & 148.85 \\
$1.5 t_0$ & 0.653 & 13.03 & -0.318 &  44.19 & 0.056 & 204.12 \\
$2.0 t_0$ & 0.649 & 13.12 & -0.303 &  35.09 & 0.042 & 264.41 \\
$2.5 t_0$ & 0.581 & 11.73 & -0.221 &  31.21 & 0.030 & 336.81 \\
$3.0 t_0$ & 0.500 &  9.61 & -0.120 &  31.34  & 0.021 & 424.09 \\
$3.5 t_0$ & 0.461 &  7.69 & -0.057 &   34.54 & 0.014 & 530.22 \\
$4.0 t_0$ & 0.457 &  6.27 & -0.026 &  40.54  & 0.009 & 660.50 \\
$4.5 t_0$ & 0.473 &  5.27 & -0.012 &  48.97  & 0.006 & 820.34 \\
$5.0 t_0$ & 0.497 &  4.58 & -0.006 &  57.87  & 0.004 & 1014.4 \\
$5.5 t_0$ & 0.522 &  4.12 & -0.003 &  61.39  & 0.003 & 1244.2 \\
$6.0 t_0$ & 0.546 &  3.82 & -0.003 &  52.30  & 0.003 & 1511.2 \\
\hline
\end{tabular}
\end{subtable}
\hspace{25mm}
\begin{subtable}{.38\textwidth}
\caption{$\eta/t_0=2.0$}
\begin{tabular}{|>{\centering\arraybackslash}p{1.4cm}|>{\centering\arraybackslash}p{1.4cm}|>{\centering\arraybackslash}p{1.4cm}|>{\centering\arraybackslash}p{1.4cm}|>{\centering\arraybackslash}p{1.4cm}|}
\hline
&\multicolumn{4}{c|}{$P(t)-P_{\text{thermal}}=\dsp{\sum_{i=1}^2} a_i \exp(-t/\tau_i)$}\\
\hline 
$U$ & $a_1$ & $\tau_1 (t_0/\hbar)$ & $a_2$ & $\tau_2 (t_0/\hbar)$ \\
\hline
$0.0 t_0$ & 0.846 & 0.63 & -0.225 & 6.54 \\
$0.5 t_0$ & 0.830 & 0.67 & -0.209 & 6.10 \\
$1.0 t_0$ & 0.818 & 0.72 & -0.200 & 5.71 \\
$1.5 t_0$ & 0.810 & 0.78 & -0.197 & 5.25 \\
$2.0 t_0$ & 0.804 & 0.85 & -0.197 & 4.73 \\
$2.5 t_0$ & 0.805 & 0.95 & -0.205 & 4.09 \\
$3.0 t_0$ & 0.882 & 1.14 & -0.294 & 3.00 \\
$3.5 t_0$ & 0.852 & 1.03 & -0.179 & 3.07 \\
$4.0 t_0$ & 0.776 & 0.90 & -0.025 & 7.67 \\
$4.5 t_0$ & 0.591 & 1.25 & -0.005 & 17.00 \\
$5.0 t_0$ & 0.551 & 1.50 & 0.009 & 7.74 \\
$5.5 t_0$ & 0.517 & 1.70 & 0.037 & 7.26 \\
$6.0 t_0$ & 0.482 & 1.89 & 0.069 & 7.30 \\
\hline
\end{tabular}
\end{subtable} 
\label{purityexp}
\end{table}

\clearpage
\bibliography{Article}

\end{document}